\documentclass{elsart}
\usepackage{color,amsmath,amssymb,graphicx}

\usepackage{mathrsfs}
\usepackage{verbatim}
\usepackage{enumerate}
\usepackage{graphics}

\newtheorem{defi}{Definition}
\newtheorem{remq}[defi]{Remark}

\newenvironment{rmq}{\begin{remq}\rm}{\end{remq}}

\newtheorem{observation}[defi]{Observation}

\newcommand{\nlr}[1]{\|#1\|_{r}}

\newcommand{\nlpsps}[1]{\|#1\|_{p+1}^{p+1}}
\newcommand{\nld}[1]{\|#1\|_2}
\newcommand{\nldd}[1]{\|#1\|_2^2}

\newcommand{\nhu}[1]{\|#1\|_{\hu}}
\newcommand{\nhud}[1]{\|#1\|_{\hu}^2}

\newcommand{\psld}[2]{(#1,#2)_{2}}

\newcommand{\dual}[2]{\left< #1,#2 \right>}

\renewcommand{\a}{\alpha}
\renewcommand{\b}{\beta}
\renewcommand{\d}{\delta}
\newcommand{\f}{\varphi}
\newcommand{\g}{\gamma}
\renewcommand{\l}{\lambda}
\newcommand{\m}{\mu}
\newcommand{\n}{\nu}

\newcommand{\s}{\sigma}

\newcommand{\w}{\omega}

\renewcommand{\L}{\Lambda}

\newcommand{\N}{\mathbb{N}}
\newcommand{\R}{\mathbb{R}}

\newcommand{\C}{\mathbb{C}}

\newcommand{\ld}{L^2(\R)}

\newcommand{\lps}{L^{p+1}(\R)}
\newcommand{\lr}{L^r(\R)}

\newcommand{\hu}{H^1(\R)}
\newcommand{\hmu}{H^{-1}(\R)}
\newcommand{\hurad}{H^1_{{\rm{rad}}}(\R)}
\newcommand{\hurc}{H^{1}(\R,\C)}
\newcommand{\hurr}{H^{1}(\R,\R)}

\newcommand{\Iwg}{I_{\w,\g}}
\newcommand{\Swg}{S_{\w,\g}}
\newcommand{\Ea}{E^{\a}}
\newcommand{\Ka}{\K^{\a}}
\newcommand{\K}{Q_\g}

\newcommand{\uo}{u_0}
\newcommand{\uu}{u_{1}}
\newcommand{\ud}{u_{2}}
\newcommand{\ua}{u^{\a}}

\newcommand{\vao}{v^{\ao}}
\newcommand{\vb}{v^{\b}}
\newcommand{\vbo}{v^{\bo}}
\newcommand{\vbot}{v^{\bot}}

\newcommand{\fwg}{\f_{\w, \g}}
\newcommand{\fwo}{\f_{\w, 0}}
\newcommand{\fwga}{\f_{\w}^{\a}}
\newcommand{\fg}{\f_{\g}}
\newcommand{\fo}{\f_{0}}
\newcommand{\fs}{f_{0}}
\newcommand{\Va}{V^\a}

\newcommand{\fa}{f^{\a}}

\newcommand{\gw}{g_{0}}

\newcommand{\Hwg}{H_{\w}^{\g}}
\newcommand{\Hg}{H^{\g}}
\newcommand{\Lug}{L_{1, \w}^{\g}}
\newcommand{\Luo}{L_{1, \w}^{0}}
\newcommand{\Ldg}{L_{2, \w}^{\g}}
\newcommand{\Lugt}{\widetilde{\Lug}}
\newcommand{\Ldgt}{\widetilde{\Ldg}}
\newcommand{\Luginf}{L_1^{\g_\infty}}
\newcommand{\Lg}{L_{1}^{\g}}
\newcommand{\Lo}{L_{1}^{0}}

\newcommand{\Bug}{B^\g_{1, \w}}
\newcommand{\Bugw}{B^\g_{1}}
\newcommand{\Bdg}{B^\g_{2, \w}}

\newcommand{\Buug}{B^\g_{1,1}}
\newcommand{\Budg}{B^\g_{1,2, \w}}
\newcommand{\Budgw}{B^\g_{1,2}}

\newcommand{\dx}{\partial_{x}}
\newcommand{\dxx}{\partial_{x}^2}
\newcommand{\dt}{\partial_{t}}
\newcommand{\dtt}{\partial_{t}^2}
\newcommand{\dw}{\partial_{\w}}

\renewcommand{\Re}{\mathrm{Re}}
\renewcommand{\Im}{\mathrm{Im}}

\renewcommand{\leq}{\leqslant}
\renewcommand{\geq}{\geqslant}

\newcommand{\goesto}{\rightarrow}
\newcommand{\goestoweak}{\rightharpoonup}

\newcommand{\Ta}{T^{\a}}
\newcommand{\Tuo}{T_{\uo}}
\newcommand{\Tue}{T_{u_\varepsilon}}
\newcommand{\Tu}{T_1}
\newcommand{\Td}{T_2}

\newcommand{\ta}{t^{\a}}
\newcommand{\va}{v^{\a}}
\newcommand{\ao}{\a_0}
\newcommand{\bo}{\b_0}
\renewcommand{\bot}{\tilde{\bo}}
\newcommand{\dM}{d_{\M}}
\newcommand{\M}{\mathscr{M}}
\newcommand{\Lginf}{\L_{\ginf}}
\newcommand{\lo}{\lambda_{0}}

\newcommand{\go}{\g_{0}}

\newcommand{\ginf}{{\g_\infty}}

\newcommand{\intr}{\int_{\R}}

\begin{document}
\begin{frontmatter}
\title{Instability of bound states of a nonlinear Schr\"odinger equation with a Dirac potential}

\author[Stefan LE COZ]{Stefan Le Coz},\ead{slecoz@univ-fcomte.fr}
\address[Stefan LE COZ]{Laboratoire de Math\'{e}matiques,
\endgraf Universit\'{e} de Franche-Comt\'{e}, 25030 Besan\c{c}on, France}

\author[Reika Fukuizumi]{Reika Fukuizumi},\ead{Reika.Fukuizumi@math.u-psud.fr}
\address[Reika Fukuizumi]{Department of Mathematics,
\endgraf Hokkaido University, Sapporo 060-0810, Japan,
\endgraf and current address: Laboratoire de Math\'{e}matiques,
\endgraf Universit\'{e} Paris Sud, 91405 Orsay, France}

\author[Fibich-Ksherim]{Gadi~Fibich, Baruch~Ksherim},\ead{fibich@tau.ac.il} 
\address[Fibich-Ksherim]{School of Mathematical Sciences, Tel Aviv University, Tel Aviv, 69978 Israel}


\author[Sivan]{Yonatan~Sivan},\ead{yonatans@post.tau.ac.il}
\address[Sivan]{School of Physics and Astronomy, Tel Aviv University, Tel Aviv, 69978 Israel}

\begin{abstract}
We study analytically and numerically the stability of the standing
waves for a nonlinear Schr\"odinger equation with a point defect and
a power type nonlinearity. A main difficulty is to compute the
number of negative eigenvalues of the linearized operator around the
standing waves, and it is overcome by a perturbation method and
continuation arguments. Among others, in the case of a repulsive
defect, we show that the standing wave solution is stable in
$\hurad$ and unstable in $\hu$ under subcritical nonlinearity.
Further we investigate the nature of instability: under critical or
supercritical nonlinear interaction, we prove the instability by
blowup in the repulsive case by showing a virial theorem and using a
minimization method involving two constraints. In the subcritical
radial case, unstable bound states cannot collapse, but rather
narrow down until they reach the stable regime (a {\em finite-width
instability}). In the non-radial repulsive case, all bound states
are unstable, and the instability is manifested by a lateral drift
away from the defect, sometimes in combination with a finite-width
instability or a blowup instability.
\end{abstract}

\begin{keyword}
Instability \sep Collapse \sep Solitary waves \sep Nonlinear waves
\sep Dirac delta \sep Lattice defects. \PACS 03.65. Ge \sep
05.45.Yv.
\end{keyword}

\end{frontmatter}

\section{Introduction}
Solitary waves are localized waves that propagate in nonlinear media
where dispersion and/or diffraction are present. They appear in
various fields of physics such as nonlinear optics, Bose-Einstein
Condensates (BEC), plasma physics, solid state physics, water waves
etc. The dynamics of solitons is modeled by the Nonlinear
Schr\"odinger equation (NLS) in the context of nonlinear optics or
the Gross-Pitaevskii (GP) equation in the context of BEC.

By now, the stability and dynamics of solitons in homogeneous media
are well understood. However, stability and dynamics of solitons in
inhomogeneous media are still a matter of intense research, both
theoretically and experimentally. Of particular interest is the NLS
equation with a linear potential (or lattice)
\begin{equation}\label{nls-lin-pot}
\left\{
\begin{array}{l}
i\dt u(t,x)  =  -\dxx u - V(x) u - |u|^{p-1} u, \\
u(0,x)  =  \uo.
\end{array}
\right.
\end{equation}
In nonlinear optics, the potential $V(x)$ describes the variation of
the linear refractive index in space. In BEC, it describes an
external potential applied to the condensate. The potential can be
localized (e.g., a single waveguide in nonlinear
optics~\cite{st,shimshon-trapping}), parabolic (e.g., a magnetic
trap in BEC~\cite{abd-review,kon-review}) or periodic (e.g., a
waveguide array or photonic crystal lattice in nonlinear
optics~\cite{segev-review}).

In the presence of a potential, a key parameter is the relative
width of the solitary wave, compared with the characteristic
length-scale of the potential. For example, in the case of a
periodic lattice, narrow solitary waves are affected, to leading
order, by the local changes of the potential near the soliton
center~\cite{fw,fsw,fsw2,narrow}, whereas wide solitary waves are
affected by the potential average over a single
period~\cite{fsw,fsw2}.

In this paper we consider a NLS/GP equation with a delta function
potential
\begin{equation}\label{nls}
\left\{
\begin{array}{l}
i\dt u(t,x)  =  -\dxx u -\g \d(x) u - |u|^{p-1}u, \\
u(0,x)  =  \uo,
\end{array}
\right.
\end{equation}
where $\g\in\R$, $1<p<+\infty$ and $(t,x)\in\R^{+}\times\R$. Here,
$\d$ is the Dirac distribution at the origin, namely,
$\dual{\d}{v}=v(0)$ for $v\in\hu$. Equation~(\ref{nls}) can be
viewed as a prototype model for the interaction of a wide soliton
with a narrow potential. The main advantage of using the
delta-function potential rather than a finite-width potential is the
existence of an explicit expression for the soliton profile. This
allows to prove results, whose proofs are considerably harder for a
general linear potential.

In nonlinear optics, Eq.~(\ref{nls}) models a soliton propagating in
a medium with a point defect~\cite{ghw,ma} or the interaction of a
wide soliton with a much narrower one in a bimodal fiber~\cite{cm}.
In BEC, this equation models the dynamics of a condensate in the
presence of an impurity of a length-scale much smaller than the
healing length. Such an impurity can be realized by a tightly
focused beam, by another spin state of the same atom or by another
alkali atom confined in an optical trap~\cite{sch2}. In contrast to
wide solitons in a periodic potential, in Eq.~(\ref{nls}) the (wide)
soliton profile is affected only by the local variation of the
potential rather than by its average. Moreover, since the potential
is localized, there is no band structure and gap solitons
characteristic of a periodic potential, see
e.g.,~\cite{efremidis-lattice-solitons}.

Equation~(\ref{nls}) was studied previously by several authors.
In~\cite{cm,ghw,hmz1,hmz2,hz,sch1,sch2}, the phenomenon of soliton
scattering by the effect of the defect was observed, namely,
interactions between the defect and the homogeneous medium soliton.
For example, varying amplitude and velocity of the soliton, they
studied how the defect is separating the soliton into two parts :
one part is transmitted past the defect, the other one is captured
at the defect. Holmer, Marzuola and Zworski \cite{hmz1,hmz2} gave
numerical simulations and theoretical arguments on this subject.
Recently, these results were observed experimentally for a single
waveguide potential~\cite{shimshon-trapping}.

In this paper, we study the stability and instability of the
standing-wave solution of (\ref{nls}) of the form $u(t,x)=e^{i\w t
}\f(x)$ where $\f$ is required to satisfy
\begin{equation}\label{snls}
\left\{
\begin{array}{l}
-\dxx \f + \w \f- \g\d(x) \f -|\f|^{p-1}\f=0,\\
\f\in\hu\setminus\{0\}.
\end{array}
\right.
\end{equation}
Stability under radial (symmetric) perturbations was studied
analytically in~\cite{ghw,foo,fj}. In this paper, we study stability
under {\em non-radial} perturbations. We also show that the
instability associated with momentum-nonconserving perturbations is
excited only for a repulsive defect ($\g < 0$), and is manifested by
a lateral movement of the wave away from the defect.

In the numerical part of this study we combine some recent ideas
such as a quantitative approach to (in)stability and
characterization of the instability type (width or drift
instability) in order to provide a systematic description of the
standing wave dynamics.
We emphasize that both our approach and results are relevant to
standing waves of the NLS~(\ref{nls-lin-pot}) with a general linear
potential, and also to NLS with a nonlinear
potential~\cite{fsw,fsw2}.

%
%

\section{Review of previous results}
{\bf Notations:} The space $L^r(\R,\C)$ will be denoted by $\lr$ and
its norm by $\nlr{\cdot}$. When $r=2$, the space $\ld$ will be
endowed with the scalar product
$$
\psld{u}{v}=\Re\intr u\bar{v}dx \, \mbox{ for } \ u,v\in\ld.
$$
The space $H^1(\R,\C)$ will be denoted by $\hu$, its norm by
$\nhu{\cdot}$ and the duality product between $\hmu$ and $\hu$ by
$\dual{\cdot}{\cdot}$. We write $\hurad$ for the space of radial (even)
functions of $\hu$ :
$$ \hurad=\{v\in \hu ;~ v(x)=v(-x), \quad x\in \R\}.$$

When $\g=0$, the set of solutions of (\ref{snls}) has been known
for a long time. In particular, modulo translation and phase,
there exists a unique positive solution, which is explicitly
known. This solution is even and is a ground state (see, for
example, \cite{bl,c,jt1} for such results). When $\g\neq 0$, an
explicit solution of (\ref{snls}) was presented in \cite{foo,ghw}
and the following was proved in \cite{fj,foo}.

\begin{prop}\label{propexistence}
Let $\w>\g^2/4.$ Then there exists a unique positive solution $\fwg$
of $(\ref{snls})$.
This solution is the unique positive minimizer of
$$
d(\w)=\left\{
\begin{array}{l}
\inf\left\{ \Swg(v) ;~ v\in\hu\setminus\{0\},~ \Iwg(v)=0 \right\}
\quad \mbox{ if } \quad \g\geq0,\\
\inf\left\{ \Swg(v) ;~ v\in\hurad\setminus\{0\},~ \Iwg(v)=0 \right\}
\quad \mbox{ if } \quad \g<0,
\end{array}
\right.
$$
where $\Swg$ and $\Iwg$ are defined for $v\in\hu$ by
\begin{eqnarray*}
\Swg(v) & = & \frac{1}{2}\nldd{\dx v}+\frac{\w}{2}\nldd{v}-\frac{\g}{2}|v(0)|^2-\frac{1}{p+1}\nlpsps{v},\\
\Iwg(v) & = & \nldd{\dx v}+\w\nldd{v}-\g|v(0)|^2-\nlpsps{v}.\\
\end{eqnarray*}
Furthermore, we have an explicit formula for $\fwg$
\begin{equation}\label{eqexplicitfwg}
\fwg(x)=\left[  \frac{(p+1)\w}{2} \mbox{{\rm{sech}}}^2\left(  \frac{(p-1)\sqrt{\w}}{2}|x|+\tanh^{-1}\left( \frac{\g}{2\sqrt{\w}} \right)  \right) \right]^{\frac{1}{p-1}}.
\end{equation}
\end{prop}

\begin{figure}[hp]
    \centerline{\scalebox{0.6}{\includegraphics{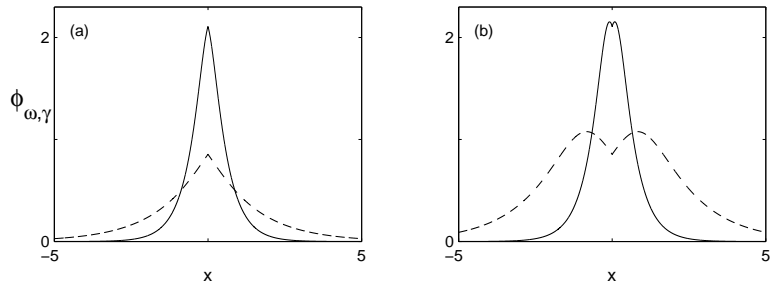}}}
 \caption{$\fwg$ as a function of~$x$ for $\omega=4$ (solid line) and $\omega=0.5$ (dashed line). (a) $\gamma=1$; (b) $\gamma=-1$. Here, $p=4$.}
  \label{fig:NLS_profiles}
\end{figure}

The dependence of
$\fwg$ on $\omega$ and $\gamma$ can be seen in Figure~\ref{fig:NLS_profiles}.
The parameter~$\omega$ affects the width and height of~$\fwg$: the
larger $\omega$ is, the narrower and higher $\fwg$
becomes, and vise versa. The sign of $\gamma$ determines the profile of
$\fwg$ near $x=0$: It has a "$\vee$" shape when $\gamma<0$,
and a "$\wedge$" shape when $\gamma>0$.

\begin{rmq}
\begin{enumerate}[(i)]
\item As it was stated in \cite[Remark 8 and Lemma 26]{fj}, the set of solutions of (\ref{snls})
$$
\{v\in\hu\setminus\{0\}\mbox{ such that }-\dxx v + \w v- \g v \d-|v|^{p-1}v=0\}
$$
is explicitly given by $\{ e^{i\theta}\fwg|\,\theta\in\R \}$.
\item There is no nontrivial solution in $\hu$ for $\w\leq \g^2/4$.
\end{enumerate}
\end{rmq}

The local well-posedness of the Cauchy problem for (\ref{nls})
is ensured by \cite[Theorem 4.6.1]{c}.
Indeed, the operator $-\dxx -\g \d$
is a self-adjoint operator on $L^2(\R)$
(see \cite[Chapter I.3.1]{aghh} and Section 2 for details). Precisely, we have
\begin{prop}\label{wpd}
For any $\uo\in\hu$, there exist $\Tuo>0$ and a unique solution $u\in\mathcal{C}([0,\Tuo),\hu)\cap\mathcal{C}^1([0,\Tuo),\hmu)$ of $(\ref{nls})$
such that $\lim_{t\uparrow \Tuo}\nld{\dx u}=+\infty$ if $\Tuo<+\infty$. Furthermore, the conservation of energy and charge hold, that is, for any $t\in[0,\Tuo)$ we have
\begin{eqnarray}
E(u(t))&=&E(\uo),\label{conservationenergy}\\
\nldd{u(t)}&=&\nldd{\uo},\label{conservationcharge}
\end{eqnarray}
where the energy $E$ is defined by
$$
E(v)=\frac{1}{2}\nldd{\dx v}-\frac{\g}{2}|v(0)|^2-\frac{1}{p+1}\nlpsps{v},
\quad \mbox{for} \quad v\in \hu.
$$
\end{prop}
(see also a verification of this proposition in \cite[Proposition 1]{foo}).

\begin{rmq}\label{rmqradial}
            From the uniqueness result of Proposition \ref{wpd} it follows that if an
initial data $u_0$ belongs to $\hurad$ then $u(t)$ also belongs to $\hurad$ for
all $t \in [0, \Tuo)$.
\end{rmq}

We consider the stability in the following sense.

\begin{defi}
Let $\f$ be a solution of $(\ref{snls})$. We say that the standing
wave $u(x,t)=e^{i\w t}\f(x)$ is \emph{(orbitally) stable} in $\hu$
(resp. $\hurad$) if for any $\varepsilon>0$ there exists $\eta >0$
with the following property : if $u_0\in \hu$ (resp. $\hurad$)
satisfies $\nhu{u_0- \f}< \eta$, then the solution $u(t)$ of
$(\ref{nls})$ with $u(0)=u_0$ exists for any $t\ge 0$ and
$$
\sup_{t \in [0,+ \infty)} \inf_{\theta \in \R}\nhu{u(t)-e^{i
\theta}\f}<\varepsilon.
$$
Otherwise, the standing wave $u(x,t)=e^{i\w t}\f(x)$ is said to be
\emph{(orbitally) unstable} in $\hu$ (resp. $\hurad$).
\end{defi}

\begin{rmq}\label{rmqstability}
With this definition and Remark \ref{rmqradial}, it is clear that
stability in $\hu$ implies stability in $\hurad$ and conversely
that instability in $\hurad$ implies instability in $\hu$.
\end{rmq}

When $\g=0$, the orbital stability for (\ref{nls}) has been
extensively studied (see \cite{bc,c,cl,ss,w} and the references
therein). In particular, from \cite{cl} we know that $e^{i \w
t}\fwo(x)$ is stable in $\hu$ for any $\w>0$ if $1<p<5$.
On the other hand,
it was shown that $e^{i \w t}\fwo(x)$ is unstable in $\hu$
for any $\w>0$ if
$p\geq 5$ (see \cite{bc} for $p>5$ and \cite{w} for $p=5$).

In \cite{ghw}, Goodman, Holmes and Weinstein focused on the
special case $p=3$, $\g>0$
and proved that the standing
wave $e^{i\w t}\fwg(x)$ is orbitally stable in $\hu$.
When $\g>0$, the orbital stability and
instability were completely studied in \cite{foo} : the standing wave $e^{i\w t}\fwg(x)$ is stable in $\hu$
for any $\w>\g^2/4$ if $1<p\leq 5$, and if $p>5$,
there exists a critical frequency $\w_1>\g^2/4$
such that $e^{i\w t}\fwg(x)$ is stable in $\hu$
for any $\w \in (\g^2/4,\w_1)$ and
unstable in $\hu$ for any $\w > \w_1$.

When $\g<0$, Fukuizumi and Jeanjean showed the following result in \cite{fj}.
\begin{prop} \label{prop:fj}
Let $\g<0$ and $\w>\g^2/4$.
\begin{itemize}
\item[(i)] If $1<p\leq3$ the standing wave $e^{i\w t}\fwg(x)$ is stable in $\hurad$.
\item[(ii)] If $3<p<5$, there exists $\w_2>\g^2/4$ such that
the standing wave $e^{i\w t}\fwg(x)$ is stable in $\hurad$
when $\w>\w_2$ and unstable in $\hu$ when $\g^2/4<\w<\w_2$.
\item[(iii)] If $p\geqslant 5$,
then the standing wave $e^{i\w t}\fwg(x)$ is unstable in $\hu$.
\end{itemize}

The critical frequency $\w_2$ is given by
\begin{eqnarray*}
&&\frac{J(\w_2)(p-5)}{p-1}
=\frac{\g}{2 \sqrt{\w_2}} \left( 1-\frac {\g^2}{4\w_2} \right)^{-(p-3)/(p-1)},\\
&&J(\w_2)=\int_{A(\w_2,\g)}^{+\infty} \mbox{sech}^{4/(p-1)}(y)dy, \quad
A(\w_2,\g)=\mbox{tanh}^{-1}\left(\frac{\g}{2\sqrt{\w_2}}\right).
\end{eqnarray*}
\end{prop}

\section{Summary of results}

The results of stability of \cite{fj} recalled in
Proposition~\ref{prop:fj} assert only on stability under radial
perturbations. Furthermore, the nature of instability is not
revealed. In this paper, we prove that there is instability in the
whole space when stability holds under radial perturbation (see
Theorem \ref{thmstability}), and that, when $p\geq 5$, the
instability established in \cite{fj} is strong instability (see
Definition \ref{def:stronginstability} and Theorem
\ref{thmblowup}).

Our first main result is the following.

\begin{thm} \label{thmstability}
Let $\g<0$ and $\w>\g^2/4$.
\begin{itemize}
\item[(i)] If $1<p\leq 3$ the standing wave $e^{i\w t}\fwg(x)$
is unstable in $\hu$.
\item[(ii)] If $3<p<5$, the standing wave $e^{i\w t}\fwg(x)$
is unstable in $\hu$ for any $\w>\w_2,$
where $\w_2$ is defined in Proposition \ref{prop:fj}.
\end{itemize}
\end{thm}

As in \cite{fj,foo}, our stability analysis relies on the abstract theory by Grillakis, Shatah and Strauss \cite{gss,gssii}
for a Hamiltonian system which is invariant under a one-parameter
group of operators. In trying to follow this approach the main point is to check the following two conditions:
\begin{enumerate}
\item The {\em slope condition}: The sign of $\dw\nldd{\fwg}$.

\item The {\em spectral condition:}  The number of negative eigenvalues of the linearized operator
$$
\Lug v = -\dxx v+\w v-\g\d v-p\fwg^{p-1}v.
$$
\end{enumerate}
We refer the reader to Section \ref{secstability} for the precise criterion and a detailed explanation on how $\Lug$ appears in this stability analysis.
Making use of the explicit form (\ref{eqexplicitfwg}) for $\fwg$, the sign of $\dw\nldd{\fwg}$ was explicitly computed in \cite{fj,foo}.

In \cite{fj}, a spectral analysis is performed to count the number of negative eigenvalues, and it is proved that
the number of negative eigenvalues of $\Lug$ in $\hurad$ is one. This spectral analysis of $\Lug$  is relying on the variational characterization of $\fwg$. However, since $\fwg$ is a minimizer only in the space of radial (even) functions $\hurad$, the result on the spectrum holds only in $\hurad$, namely for even eigenfunctions. Therefore the number of negatives eigenvalues is known only for $\Lug$ considered in $\hurad$. With this approach,
it is not possible to see whether other negative eigenvalues appear when the problem is considered on the whole space $\hu$.

To overcome this difficulty, we develop a perturbation method.
In the case $\g=0$, the spectrum of $\Luo$ is well known
by the work of Weinstein \cite{w2} (see Lemma \ref{lemweinstein}) : there is only one negative eigenvalue, and $0$ is a simple isolated eigenvalue (to see that, one proves that the kernel of $\Luo$ is spanned by
$\dx \fwo$, that $\dx \fwo$ has only one zero, and apply the Sturm Oscillation Theorem). When $\g$ is small, $\Lug$ can be considered as a holomorphic perturbation of $\Luo$. Using the theory of holomorphic perturbations for linear operators, we prove that the spectrum of $\Lug$ depends holomorphically on the spectrum of $\Luo$ (see Lemma \ref{lemanalyticity}). Then the use of Taylor expansion for the second eigenvalue of $\Lug$ allows us
to get the sign of the second eigenvalue when $\g$ is small (see Lemma \ref{lemvpneg}). A continuity argument combined with the fact that if $\g\neq 0$ the nullspace of $\Lug$ is zero extends the result to all $\g\in\R$ (see the proof of Lemma \ref{lem:propnegev}). See subsection \ref{subsecpropnHwg} for details. We will see that there are two negative eigenvalues of $\Lug$ in $\hu$
if $\g<0$.

\begin{rmq}
\begin{enumerate}[(i)]
\item
Our method can be applied as well in $\hu$ or in $\hurad$, and for $\g$ negative or positive (see subsections \ref{subsec:positivecase} and \ref{subsecradialcase}). Thus we can give another proof of the result of \cite{foo} in the case $\g>0$ and of Proposition \ref{prop:fj}.
\item
The study of the spectrum of linearized operators
is often a central point when one wants to use the abstract theory of \cite{gss,gssii}.
See \cite{fsw,gh1,gh2,gs,jl} among many others for related results.
\end{enumerate}
\end{rmq}

The results of instability given
in Theorem \ref{thmstability} and Proposition \ref{prop:fj}
say only that a certain solution which
starts close to $\fwg$ will exit from a tubular neighborhood
of the orbit of the standing wave in finite time.
However, as this might be of
importance for the applications, we want to understand further the
nature of instability. For that, we recall the concept of strong
instability.

\begin{defi}\label{def:stronginstability}
A standing wave $e^{i\w t}\f(x)$ of $(\ref{nls})$ is said to be
\emph{strongly unstable in $\hu$ } if for any $\varepsilon>0$
there exist $u_\varepsilon\in H^1(\R)$ with
$\nhu{u_\varepsilon-\f}<\varepsilon$ and $\Tue<+\infty$ such that
$\lim_{t\uparrow \Tue}\nld{\dx u(t)}=+\infty$, where $u(t)$  is
the solution of $(\ref{nls})$ with $u(0)=u_\varepsilon$.
\end{defi}

Our second main result is the following.

\begin{thm}\label{thmblowup}
Let $\g\leq 0$, $\w>\g^2/4$ and $p\geqslant 5$.
Then the standing wave $e^{i\w t}\fwg(x)$
is strongly unstable in $\hu$.
\end{thm}

Whether the perturbed standing wave blows-up or not depends on the
perturbation. Indeed, in Remark~\ref{globalinstability} we define
an invariant set of solutions and show that if we consider an
initial data in this set, then  the solution exists globally even
when the standing wave $e^{i\w t}\fwg(x)$ is strongly unstable.

We also point out that when $1<p<5$, it is easy to prove using the conservation laws and
Gagliardo-Nirenberg inequality that the Cauchy problem in $\hu$
associated with~(\ref{nls}) is globally well posed. Accordingly,
even if the standing wave may be unstable
when $1<p<5$ (see Theorem \ref{thmstability}), a strong instability cannot occur.

As in \cite{bc,w}, which deal with the classical case $\g=0$,
we use the virial identity for the proof of Theorem \ref{thmblowup}.
However, even if the formal calculations are similar
to those of the case $\g=0$, a rigorous
proof of the virial theorem does not immediately follow from the approximation
by regular solutions (e.g. see \cite[Proposition 6.4.2]{c}, or \cite{gv}).
Indeed, the argument in \cite{c} relies on the $H^2(\R)$ regularity of the
solutions of (\ref{nls}). Because of the defect term, we do not know if this $H^2(\R)$ regularity still hold when $\g\neq 0$. Thus we need another approach. We approximate the solutions of (\ref{nls})
by solutions of the same equation where the defect is approximated by a Gaussian potential for which it is easy to have the virial
theorem. Then we pass to the limit in the virial identity to obtain :

\begin{prop}\label{thmvirial}
Let $\uo\in\hu$ such that $x\uo\in\ld$ and $u(t)$ be the solution
of $(\ref{nls})$. Then the function $f:t\mapsto\nldd{xu(t)}$ is
$\mathcal{C}^2$ and
\begin{eqnarray}
\dt f(t)  & = & 4\Im\intr \bar{u}x\dx u dx, \label{eqvirial1}\\
\dtt f(t) & = & 8\K(u(t)),\label{eqvirial2}
\end{eqnarray}
where $\K$ is defined for $v\in\hu$ by
$$
\K(v)=\nldd{\dx v}-\frac{\g}{2}|v(0)|^2
-\frac{p-1}{2(p+1)}\nlpsps{v}.
$$
\end{prop}

Even if we  benefit from the virial identity,
the proofs given in \cite{bc,w} for the case $\g=0$
do not apply to the case $\g<0$.
For example, the method of
Weinstein \cite{w} in the case $p=5$ requires
in a crucial way an equality between
$2E$ and $Q$ which does not hold anymore
when $\g<0$. Moreover, the heart of the
proof of \cite{bc} consists in minimizing the functional $\Swg$ on the
constraint $\K(v)=0$, but the standard variational methods to prove such results
are not so easily applied to the case of $\g\neq 0$.
To get over these difficulties we introduce an approach based on a
 minimization problem involving two constraints.
Using this minimization problem, we identify some invariant properties
under the flow of (\ref{nls}). The combination with these invariant properties
and the conservation of energy and charge allows us to prove strong instability.
We mention that some related techniques have been
introduced in \cite{l1,l2,lww,ot,z}.
In conclusion, we can give a simpler method
to prove Theorem \ref{thmblowup} than that of \cite{bc}
even though we have a term of delta potential.

\begin{rmq}
 The case $\g<0$, $\w=\w_2$ and $3<p<5$ cannot
be treated with our approach and is left open (see Remark \ref{rmq:opencase}).
In light of Theorem~\ref{thmstability}, we believe that the standing wave is unstable
in this case, at least in $\hu$ (see also \cite[Remark 12]{fj}). When $\g>0$, the case $\w=\w_1$ and $p>5$ is also open (see \cite[Remark 1.5]{foo}).
\end{rmq}

Let us summarize the  previously known and our  new rigorous results
on stability in~(\ref{nls}):
\begin{enumerate}[(i)]
   \item  For both positive and negative $\gamma$, there is always only one negative eigenvalue of linearized operator in~$\hurad$ (\cite{fj}, subsection 2.5). Hence,
the standing wave is stable in $\hurad$ if the slope  is positive, and unstable if the slope is negative.
   \item $\gamma>0$. In this case the number of the negative eigenvalues
of linearized operator is always one in $\hu$. Stability is determined by
  the slope condition, and the standing wave is stable in $\hurad$  if and only if it is stable in  $\hu$. Specifically (\cite{fj,foo},  subsection 2.4),
\begin{enumerate}
   \item $1<p\le 5$: Stability in $\hu$ for any $\omega> \gamma^2/4$.
   \item $5<p$: Stability in $\hu$ for $\gamma^2/4< \omega < \omega_1$,
instability in $\hurad$ for  $\omega > \omega_1$.
\end{enumerate}
   \item $\gamma<0$.
In this case the number of negative eigenvalues is always two
(Lemma \ref{lem:propnegev})
and all standing waves are unstable in  $\hu$ (Theorem~\ref{thmstability}
and Theorem~\ref{thmblowup}).
Stability in  $\hurad$ is determined by
  the slope condition and is as follows (\cite{fj}):
\begin{enumerate}
   \item $1<p \le 3$: Stability in  $\hurad$ for any $\omega> \gamma^2/4$.
   \item $3<p < 5$: Stability in $\hurad$  for $\omega > \omega_2$,
instability in $\hurad$ for  $\gamma^2/4 <\omega < \omega_2$.
   \item $5 \le p$: Strong instability in $\hurad$ (and in  $\hu$)
for any $\gamma^2/4<\omega$ (Theorem \ref{thmblowup}).
\end{enumerate}
\end{enumerate}

    There are, however, several important questions which are still open, and which we
   explore using numerical simulations. Our simulations suggest the following:
\begin{enumerate}[(i)]
   \item Although an attractive defect ($\gamma>0$) stabilizes
the standing waves in the critical case ($p=5$),
their stability is weaker than in the subcritical case,
in particular for $0 < \gamma \ll 1$.
   \item  Theorem \ref{thmblowup} shows that instability occurs by blowup
  when $\g <0$ and $p \geq 5$. In all other cases, however, it remains to understand the nature of
instability. Our simulations suggest the following:

\begin{enumerate}
   \item When $\g>0, \ p>5,$ and $\w > \w_1$, instability can occur by blowup.
   \item  When $\g<0, \ 3<p<5$, and $ \gamma^2/4< \w < \w_2$, the instability in $\hurad$ is a {\em finite-width
instability}, i.e., the solution initially narrows down along a curve
$\phi_{\omega*(t), \gamma}$, where $\omega^*(t)$ can be defined by
the relation
\begin{eqnarray*}
    \max_{x} \phi_{\omega^{*}(t),\gamma}(x) = \max_{x} |u(x,t)|.
    \label{eq:omega_min_condition}
\end{eqnarray*}
As the solution narrows down, $\omega^*(t)$ increases and crosses
from the unstable region $\omega <\omega_{2}$
to the stable region $\omega >\omega_{2}$. Subsequently, collapse is arrested at some finite width.
   \item When $\gamma<0$, the standing waves undergo a {\em drift instability}, away from the
          (repulsive) defect, sometimes in combination with finite-width or blowup instability. Specifically,
\begin{itemize}
    \item[(c.i)] When $1<p \le 3$ and when $3<p <5 $ and $\w >\w_2$ (i.e., when the standing waves are
  stable in $\hurad$),
  the standing waves undergo a {\em drift instability}.
    \item[(c.ii)] When $3<p<5$ and $\gamma^2/4<\w < \w_2$,   the instability in $\hu$ is a combination of a
   drift instability and a finite-width instability.
    \item[(c.iii)] When $p \ge 5$, the instability in $\hu$ is a combination of a drift instability and a blowup instability.
\end{itemize}
\end{enumerate}
   \item Although  when $p=5$ and $\gamma>0$, and when
$ p>5$, $\g>0, $ and $\gamma^2/4<\w < \w_1$ the standing wave is stable, it can collapse under a sufficiently large perturbation.
\end{enumerate}

We note that all of the above holds, more generally, for NLS
equations with a nonlinear potential~\cite{fsw,fsw2} and for narrow
solitons of a linear potential~\cite{narrow}.

The rest of the paper is organized as follows. In Section
\ref{secstability}, we prove Theorem \ref{thmstability} and explain
how our method allows us to recover the results of \cite{fj,foo}. In
Section~\ref{secblowup}, we establish Theorem \ref{thmblowup}.
Numerical results are given in Section~\ref{sec:numerics}.

Throughout the paper the letter $C$ will denote various
positive constants whose exact values may change from line to line but are not
essential to the analysis of the problem.

\textbf{Acknowledgements.}

The authors are grateful to Louis Jeanjean for fruitful discussions
and helpful advice. Reika Fukuizumi would like to thank Shin-ichi
Shirai and Cl\'{e}ment Gallo for useful discussions particularly
about Section 2. {Stefan Le Coz wishes to thank Mariana
H\u{a}r\u{a}gu\c{s} for fruitful discussions. The research of Gadi
Fibich, Baruch Ksherim and Yonatan Sivan was partially supported by
grant 2006-262 from the United States-Israel Binational Science
Foundation (BSF), Jerusalem, Israel.}

\section{Instability with respect to non-radial perturbations}\label{secstability}

We use the general theory of Grillakis, Shatah and Strauss \cite{gssii}
to prove Theorem \ref{thmstability}.

First, we explain how we derive a criterion for stability or
instability for our case from the theory of Grillakis, Shatah and
Strauss. In our case, it is clear that \textit{Assumption 1} and
\textit{Assumption 2} of \cite{gssii} are satisfied. The last
assumption, \textit{Assumption 3}, should be checked. We consider
the sesquilinear form $\Swg''(\fwg):\hu\times\hu\goesto\C$ as a
linear operator $\Hwg:\hu\goesto\hmu$. The spectrum of $\Hwg$ is the
set $\{ \l\in\C\mbox{ such that }\Hwg-\l I\mbox{ is not invertible}
\}$, where $I$ denote the usual $\hu-\hmu$ isomorphism, and we
denote by $n(\Hwg)$ the number of negative eigenvalues of $\Hwg$.
Having established the assumptions of \cite{gssii}, the next
proposition follows from \cite[Instability Theorem and Stability
Theorem]{gssii}.

\begin{prop}\label{prop:criterion}
\begin{enumerate}[$(1)$]
\item The standing wave
$e^{i\w_0 t} \f_{\w_0, \g}(x)$ is unstable
if the integer $(n(H_{\w_0}^{\g})-p(d''(\w_0))$ is odd,
where
$$
p(d''(\w_0))=\left\{
\begin{array}{l}
1 \quad \mbox{if} \quad \partial_{\w}\nldd{\fwg}>0
\quad \mbox{at} \quad \w=\w_0,  \\
0 \quad \mbox{if} \quad \partial_{\w}\nldd{\fwg}<0
\quad \mbox{at} \quad \w=\w_0.
\end{array}
\right.
$$
\item The standing wave
$e^{i\w_0 t} \f_{\w_0, \g}(x)$ is stable
if $(n(H_{\w_0}^{\g})-p(d''(\w_0))=0$.
\end{enumerate}
\end{prop}

Let us now consider the case $\g<0$. It was proved in \cite{fj} that
\begin{lem}\label{lem:signdwgneg}
Let $\g<0$ and $\w>\g^2/4$. We have :
\begin{itemize}
\item[$($i$)$] If $1<p\leq3$  and $\w>\g^2/4$  then  $\dw\nldd{\fwg}>0$,
\item[$($ii$)$] If $3<p<5$  and $\w>\w_2$  then  $\dw\nldd{\fwg}>0$,
\item[$($iii$)$] If $3<p<5$  and $\g^2/4<\w<\w_2$  then  $\dw\nldd{\fwg}<0$,
\item[$($iv$)$] If $p\geq 5$  and $\w>\g^2/4$  then  $\dw\nldd{\fwg}<0$.
\end{itemize}
\end{lem}

Thus Theorem \ref{thmstability} follows from Proposition \ref{prop:criterion}, Lemma \ref{lem:signdwgneg} and

\begin{lem}\label{propnHwg}
If $\g<0$, then $n(\Hwg)=2$.
\end{lem}

\begin{rmq}\label{rmq:opencase}
\begin{enumerate}
\item Let $\g<0$.
In the cases $3<p<5$ and $\w<\w_2$ or $p\ge 5$ it was proved in
\cite{fj} that $\dw\nldd{\fwg}<0$. From Lemma \ref{propnHwg}, we
know that the number of negative eigenvalues of $\Hwg$ is
$n(\Hwg)=2$ when $\Hwg$ is considered on the whole space $\hu$.
Therefore $n(\Hwg)-p(d''(\w))=2$ and this corresponds to a case
where the assumption of \cite{gssii} may not be applied. However,
if we consider $\Hwg$ in $\hurad$, then it follows from \cite{fj}
that $n(\Hwg)=1$, thus $n(\Hwg)-p(d''(\w))=1$. Then, we can apply
Proposition \ref{prop:criterion} to this case and it allows us to
conclude instability in $\hurad$ (as it was done in \cite{fj}).
But, with Remark \ref{rmqstability}, we can conclude that
instability holds on the whole space $\hu$.
\item Note that the case $\w=\w_2$
corresponds to $\partial_{\w}\nldd{\fwg}=0$ ($3<p<5$) and will not be treated here.
In view of Theorem \ref{thmstability},
we believe that the standing wave is unstable in this case, at least in $\hu$.
\end{enumerate}
\end{rmq}

We divide the rest of this section into four parts. In subsection
\ref{subsecsettings} we introduce the general setting to perform our
proof, and we study whether \textit{Assumption 3} of \cite{gssii} is
satisfied. Lemma \ref{propnHwg} will be proved in subsection
\ref{subsecpropnHwg}. Finally, we discuss the positive case and the
radial case in subsections \ref{subsec:positivecase} and
\ref{subsecradialcase}.

\subsection{Setting for the spectral problem}
\label{subsecsettings}

To express $\Hwg$,
it is convenient to split $u$ in real and imaginary part :
for $u\in\hurc$ we write $u=\uu+i\ud$
where $\uu=\Re(u)\in\hurr$ and $\ud=\Im(u)\in\hurr$. Now we set
$$
\Hwg u=\Lug \uu+i\Ldg\ud
$$
where the operators $\Lug,\Ldg:\hurr\goesto\hmu$ are defined for $v\in\hu$ by
\begin{eqnarray*}
\Lug v& = &-\dxx v+\w v-\g v \d-p\fwg^{p-1}v,\\
\Ldg v& = &-\dxx v+\w v-\g v \d -\fwg^{p-1}v.
\end{eqnarray*}
When we will work with $\Lug,\Ldg$, the functions considered will be understood to be real
valued.

For the spectral study of $\Hwg$, it is convenient to view $\Hwg$ as an unbounded operator on $\ld$, thus we rewrite our spectral problem in this setting. First, we redefine the two operators $\Lug$ and $\Ldg$ as unbounded operators on $\ld$. We begin by considering the bilinear forms on $\hu$ associated with $\Lug$ and $\Ldg$ by setting for $v,w\in\hu$
$$
\Bug(v,w):=\dual{\Lug v}{w}\mbox{ and }\Bdg(v,w):=\dual{\Ldg v}{w},
$$
which are explicitly given by
\begin{equation}\label{eqBugBdg}
\begin{array}{rcl}
\Bug(v,w)&=&\intr{\dx v\dx w dx}\!+\!\w\!\intr{vw dx}\!-\!\g v(0)w(0)\!-\!\intr p\fwg^{p-1}vwdx,\\
\Bdg(v,w)&=&\intr{\dx v\dx w dx}\!+\!\w\!\intr{vw dx}\!-\!\g v(0)w(0)\!-\!\intr \fwg^{p-1}vwdx.
\end{array}
\end{equation}

Let us now consider $\Bug$ and $\Bdg$ as bilinear forms on $\ld$ with domain $D(\Bug)=D(\Bdg):=\hu$. It is clear that theses forms are bounded from below and closed. Then the theory of representation of forms by operators (see \cite[VI.\textsection 2.1]{k}) implies that we define two self-adjoint operators $\Lugt:D(\Lugt)\subset\ld\goesto\ld$ and $\Ldgt:D(\Ldgt)\subset\ld\goesto\ld$ by setting
\begin{eqnarray*}
D(\Lugt)&:=&\{ v\in\hu|\exists w\in\ld \mbox{ s.t. }\forall z\in\hu,\,\Bug(v,z)=\psld{w}{z} \},\\
D(\Ldgt)&:=&\{ v\in\hu|\exists w\in\ld \mbox{ s.t. }\forall z\in\hu,\,\Bdg(v,z)=\psld{w}{z} \}.
\end{eqnarray*}
and setting for $v\in D(\Lugt)$ (resp. $v\in D(\Ldgt)$) that $\Lugt v:=w$ (resp. $\Ldgt v:=w$), where $w$ is the (unique) function of $\ld$ which satisfies $\Bug(v,z)=\psld{w}{z}$ (resp. $\Bdg(v,z)=\psld{w}{z}$) for all $z\in\hu$.

For notational simplicity, we drop the tilde over $\Lugt$ and $\Ldgt$.
It turns out that we are able to describe explicitly $\Lug$ and $\Ldg$.

\begin{lem}\label{lemmadefLug}
The domain of $\Lug$ and of $\Ldg$ in $\ld$ is
$$
D_\g =\{v\in\hu\cap H^2(\R\setminus\{0\});~\dx v(0^+)-\dx v(0^-)
=-\g v(0)\}
$$
and for $v\in D_\g$ the operators are given by
\begin{equation}\label{Lug}
\begin{array}{rcl}
\Lug v& = &-\dxx v+\w v-p\fwg^{p-1}v, \\
\Ldg v& = &-\dxx v+\w v-\fwg^{p-1}v.
\end{array}
\end{equation}
\end{lem}

The proof of this lemma is given in Appendix~\ref{ap:lemmadefLug}.
We conclude this subsection mentioning some basic properties of the
spectrum of $\Hwg$. Precisely, to check \cite[\textit{Assumption
3}]{gssii} is equivalent to check the following lemma.

\begin{lem}\label{lemma28}
Let $\g\in\R\setminus\{0\}$ and $\w>\g^2/4$.
\begin{itemize}
\item[$(\mathrm{i})$] The operator $\Hwg$ has only
a finite number of negative eigenvalues,
\item[$(\mathrm{ii})$] The kernel of $\Hwg$ is $\mbox{\rm span}\{i\fwg\}$,
\item[$(\mathrm{iii})$] The rest of the spectrum of $\Hwg$ is positive
and bounded away from $0$.
\end{itemize}
\end{lem}

Our proof of Lemma \ref{lemma28} borrows some elements of \cite{fj}.
In particular, (ii) in Lemma \ref{lemma28} is shown in \cite[Lemma
28 and Lemma 31]{fj}. For the sake of completeness, we provide a
proof in Appendix~\ref{ap:28-analyticityLug}.

\subsection{Count of the number of negative eigenvalues}
\label{subsecpropnHwg}

In this subsection, we prove Lemma \ref{propnHwg}. First, we remark that, as it was shown in the proof of Lemma \ref{lemma28}, $0$ is the first eigenvalue of $\Ldg$. Thus $n(\Hwg)=n(\Lug)$, where $n(\Lug)$ is the number of negative eigenvalues of $\Lug$. Therefore, Lemma \ref{propnHwg} follows from
\begin{lem}\label{lem:propnegev}
Let $\g<0$ and $\w>\g^2/4$. Then $n(\Lug)=2$.
\end{lem}

Our proof of Lemma \ref{lem:propnegev} is divided in two steps.
First, we use a perturbative approach to prove that, if $\g$ is close to $0$ and negative,
$\Lug$ has two negative eigenvalues (Lemma \ref{lemvpneg}).
To do this, we have to ensure that the eigenvalues
and the eigenvectors are regular enough with respect to $\g$ (Lemma \ref{lemanalyticity})
to make use of Taylor formula.
It follows from the use of the analytic perturbation theory of operators (see \cite{k,rs}).
The second step consists
in extending the result of the first step to any values of $\g<0$. Our argument relies on the continuity of the spectral projections with respect to $\g$ and it is crucial, as it was proved in Lemma \ref{lemma28}, that $0$ can not be an eigenvalue of $\Lug$ (see \cite{gh1,gh2} for related arguments).

We fix $\w> \g^2/4$. For the sake of simplicity we denote $\Lug$ by
$\Lg$ and $\fwg$ by $\fg$, and so on in this section 2. The
following lemma verifies the holomorphicity of the operator $\Lug$,
see proof in Appendix~\ref{ap:28-analyticityLug}.

\begin{lem}\label{lemanalyticityLug}
As a function of $\g$, $(\Lg)$ is a real-holomorphic family
of self-adjoint operators (of type (B) in the sense of Kato).
\end{lem}

The following classical result of Weinstein \cite{w2} gives a precise description of the spectrum
of the operator we want to perturb.

\begin{lem}\label{lemweinstein}
The operator $\Lo$ has exactly one negative simple isolated first eigenvalue. The second eigenvalue is $0$, and
it is simple and isolated. The nullspace is ${\rm{span}}\{\dx\fo\}$,
and the rest of the spectrum is positive.
\end{lem}

Combining Lemma \ref{lemanalyticityLug} and Lemma \ref{lemweinstein},
we can apply the theory of analytic perturbations for linear operators
(see \cite[VII.\textsection 1.3]{k}) to get the following lemma.
Actually, the perturbed eigenvalues are holomorphic since they are simple.

\begin{lem}\label{lemanalyticity}
There exist $\go>0$ and two functions $\l:(-\go,\go)\mapsto\R$
and $f:(-\go,\go)\mapsto\ld$ such that
\begin{itemize}
\item[$(\mathrm{i})$] $\l(0)=0$ and $f(0)=\dx\fo$,
\item[$(\mathrm{ii})$] For all $\g\in(-\go,\go)$, $\l(\g)$
is the simple isolated second eigenvalue of $\Lg$
and $f(\g)$ is an associated eigenvector,
\item[$(\mathrm{iii})$] $\l(\g)$ and $f(\g)$ are holomorphic in $(-\go,\go)$.
\end{itemize}
Furthermore, $\go>0$ can be chosen small enough to ensure that,
expect the two first eigenvalues, the spectrum of $\Lg$ is positive.
\end{lem}

Now we investigate how the perturbed second eigenvalue moves
depending on the sign of $\g$.

\begin{lem}\label{lemvpneg}
There exists $0<\g_1<\go$ such that $\l(\g)<0$
for any $-\g_1<\g<0$ and $\l(\g)>0$ for any $0<\g<\g_1$.
\end{lem}

{\it Proof of Lemma \ref{lemvpneg}.}
We develop the functions $\l(\g)$ and $f(\g)$ of Lemma \ref{lemanalyticity}.
There exist $\lo\in\R$ and $\fs\in\ld$
such that for $\g$ close to $0$ we have
\begin{eqnarray}
\l(\g) & = & \g\lo+O(\g^2),\label{eqdev1}\\
f(\g) & = & \dx\fo+\g\fs+O(\g^2).\label{eqdev2}
\end{eqnarray}
            From the explicit expression (\ref{eqexplicitfwg}) of $\fg$, we deduce that
there exists $\gw\in\hu$ such that for $\g$ close to $0$ we have
\begin{equation}\label{eqdev3}
\fg=\fo+\g\gw+O(\g^2).
\end{equation}
Furthermore, using (\ref{eqdev3}) to substitute into (\ref{snls})
and differentiating (\ref{snls}) with respect to $\g$,
we obtain
\begin{equation}\label{Luogw}
\dual{\Lo\gw}{\psi}= \fo(0) \psi(0)+O(\g),
\end{equation}
for any $\psi \in H^1(\R)$.

To develop $\lo$ with respect to $\g$,
we
compute $\psld{\Lg f(\g)}{\dx\fo}$ in two different ways.

On one hand, using $\Lg f(\g)= \l(\g) f(\g)$, (\ref{eqdev1}) and (\ref{eqdev2}) lead us to
\begin{equation}\label{eqway1}
\psld{\Lg f(\g)}{\dx\fo}=\lo\g\nldd{\dx\fo}+O(\g^2).
\end{equation}

On the other hand, since $\Lg$ is self-adjoint, we get
\begin{equation}\label{eqway2.1}
\psld{\Lg f(\g)}{\dx\fo}=\psld{f(\g)}{\Lg\dx\fo}.
\end{equation}
Here we note that $\dx\fo\in D(\Lg)$ : indeed,
$\dx\fo\in H^2(\R)$ and $\dx\fo(0)=0$.
We compute the right hand side of (\ref{eqway2.1}).
We use (\ref{Lug}), $\Lo\dx\fo=0$, and (\ref{eqdev3})
to obtain
\begin{eqnarray}
\Lg\dx\fo&=&p(\fo^{p-1}-\fg^{p-1})\dx\fo,\nonumber\\
&=&-\g p(p-1)\fo^{p-2} g_0 \dx\fo+O(\g^2).\label{eqway2.5}
\end{eqnarray}
Hence, it follows from (\ref{eqdev2}) that
\begin{equation}\label{eqway2.2}
\psld{\Lg f(\g)}{\dx\fo}=-\psld{\dx\fo}{\g\gw p(p-1)\fo^{p-2}\dx\fo}+O(\g^2).
\end{equation}

Now, as it was remarked in \cite[Lemma
28]{fsw}, it is easy to see that using (\ref{snls}) with $\g=0$ we get

\begin{equation}\label{eqway2.6}
\Lo(\fo-\fo^{p-1})=p(p-1)\fo^{p-2}\dx\fo^2,
\end{equation}
which combined with (\ref{eqway2.2}) gives
\begin{equation}\label{eqway2.3}
\psld{\Lg f(\g)}{\dx\fo}=-\g\dual{\Lo\gw}{\fo-\fo^{p}}+O(\g^2).
\end{equation}
Finally, with (\ref{Luogw}) we obtain from (\ref{eqway2.3})
\begin{equation}\label{eqway2.4}
\psld{\Lg f(\g)}{\dx\fo}=-\g(\fo(0)^2-\fo(0)^{p+1})+O(\g^2).
\end{equation}

Combining (\ref{eqway2.4}) and (\ref{eqway1}) we obtain
$$
\lo=-\frac{\fo(0)^2-\fo(0)^{p+1}}{\nldd{\dx\fo}}+O(\g).
$$
It follows that $\lo$ is positive for sufficiently small $|\g|$,
which in view of (\ref{eqdev1}) ends the proof.
\hfill\qed

We are now in position to prove Lemma \ref{lem:propnegev}.

{\it Proof of Lemma \ref{lem:propnegev}.}
Let $\ginf$ be defined by
$$
\ginf=\inf\{\tilde{\g}<0;~\Lg\mbox{ has exactly two negative eigenvalues for all}~ \g\in(\tilde{\g},0] \}.
$$
            From Lemma \ref{lemvpneg}, we know that $\ginf$
is well defined and $\ginf\in[-\infty,0)$. Arguing by contradiction, we suppose $\ginf>-\infty$.

Let $N$ be the number of negative eigenvalues of $\Luginf$. Denote the first eigenvalue
of $\Luginf$ by $\Lginf$. Let $\Gamma$ be defined by
\begin{equation*}
\Gamma=\{z\in\C;~ z=z_1+iz_2,~ (z_1,z_2) \in [-b,0] \times [-a,a],
\mbox{ for some } a>0, b>|\Lginf|\}.
\end{equation*}
            From Lemma \ref{lemma28}, we know that $\Luginf$ does not admit zero as
eigenvalue. Thus $\Gamma$ define a contour in $\C$ of the segment
$[\Lginf,0]$ containing no positive part of the spectrum of $\Luginf$, and
without any intersection with the spectrum  of $\Luginf$. It is easily seen
(for example, along the lines of the proof of \cite[Theorem VII-1.7]{k}) that
there exists a small $\g_*>0$ such that for any $\g\in[\ginf-\g_*,\ginf+\g_*]$,
we can define a holomorphic projection on the negative part of the spectrum of
$\Lg$ contained in $\Gamma$ by
$$\Pi(\g)=\frac{-1}{2\pi i} \int_\Gamma (\Lg-z)^{-1}dz.$$
Let us insist on the fact that we can choose $\Gamma$ independently
of the parameter $\gamma$ because $0$ is not an eigenvalue of $\Lg$
for all $\gamma$.

Since $\Pi$ is holomorphic, $\Pi$ is continuous in $\g$, then by a classical connectedness argument
(for example, see \cite[Lemma I-4.10]{k}), we know that
$\dim(\mbox{Ran } \Pi(\g))=N$ for any $\g\in[\ginf-\g_*,\ginf+\g_*]$.
Furthermore, $N$ is exactly the number of negative eigenvalues of $\Lg$
when $\g\in[\ginf-\g_*,\ginf+\g_*]$ :
indeed, if $\Lg$ has a negative eigenvalue outside of $\Gamma$
it suffice to enlarge $\Gamma$ (i.e., enlarge $b$) until
it contains this eigenvalue
to
raise a contradiction since then $L_1^{\ginf}$ would have, at least, $N+1$
eigenvalues. Now by the definition of $\g_{\infty}$, $L_1^{\ginf+\g_*}$ has two
negative eigenvalues and thus we see that $L_1^{\g}$ has two negative
eigenvalues for all $\g \in [\g_{\infty}- \g^*, 0[$ contradicting the
definition of $\g_{\infty}. $

Therefore
$\ginf=-\infty$.
\hfill\qed

\begin{rmq}
In \cite[Lemma 32]{fj}, the authors proved that
there are \emph{at most} two negative eigenvalues of $\Lg$
in $\hu$ using variational methods.
In our present proof, we can directly show
that there are exactly two negative eigenvalues without
such variational techniques.
\end{rmq}

\subsection{The case $\gamma>0$}\label{subsec:positivecase}
The proof of Lemma \ref{lem:propnegev} can be easily adapted to the case $\g>0$, and with Lemma \ref{lemvpneg} we can infer
that $\Lg$ has only one simple negative eigenvalue when $\g>0$. Since $n(\Hg)=n(\Lg)$, it follows that (in the following Lemmas
\ref{lem:positivecase}, \ref{lem:signdwgpos} and Proposition \ref{prop:theorem1foo},  there is no omission of parameter $\omega$ to understand the dependence clearly)
\begin{lem}\label{lem:positivecase}
Let $\g>0$ and $\w>\g^2/4$. Then the operator $\Hwg$ has only one negative eigenvalue, that is $n(\Hwg)=1$.
\end{lem}
When $\g>0$, the sign of $\dw\nldd{\fwg}$ was computed in \cite{foo}. Precisely :
\begin{lem}\label{lem:signdwgpos}
Let $\g>0$ and $\w>\g^2/4$. We have :
\begin{itemize}
\item[$($i$)$] If $1<p\leq5$  and $\w>\g^2/4$  then  $\dw\nldd{\fwg}>0$,
\item[$($ii$)$] If $p> 5$  and $\g^2/4<\w<\w_1$  then  $\dw\nldd{\fwg}>0$,
\item[$($iii$)$] If $p> 5$  and $\w>\w_1$  then  $\dw\nldd{\fwg}<0$.
\end{itemize}
Here $\w_1$ is defined as
follows:
\begin{eqnarray*}
&& \frac{p-5}{p-1} J(\w_1)=\frac{\g}{2\sqrt{\w_1}}
\left(1-\frac{\g^2}{4\w_1} \right)^{-(p-3)/(p-1)},\\
&& J(\w_1) =\int_{A(\w_1, \g)}^{\infty}
\mathrm{sech}^{4/(p-1)} ydy, \quad A(\w_1, \g) =
\mathrm{tanh}^{-1} \left(\frac{\g}{2\sqrt{\w_1}} \right).
\end{eqnarray*}
\end{lem}
Then, using Lemma \ref{lem:positivecase}, Lemma \ref{lem:signdwgpos} and Proposition \ref{prop:criterion}, we can give an alternative proof of \cite[Theorem 1]{foo} (see also \cite[Remark 33]{fj}). Precisely, we obtain\,:
\begin{prop}\label{prop:theorem1foo}
Let $\g>0$ and $\w>\g^2/4$.
\begin{itemize}
\item[$($i$)$] Let $1<p\leq 5$.
Then $e^{i\w t}\fwg(x)$ is stable in $\hu$ for any
$\w\in (\g^2/4,+\infty)$.
\item[$($ii$)$] Let $p> 5$.
Then $e^{i\w t}\fwg(x)$ is stable in $\hu$ for any
$\w\in (\g^2/4, \w_1)$, and unstable in $\hu$ for any
$\w\in (\w_1, +\infty)$.
\end{itemize}
\end{prop}

\subsection{The radial case}\label{subsecradialcase}
Before we start to discuss the stability in the radial case,
we mention the following remarkable fact.

\begin{lem} \label{lem:impaire}
The function $f(\g)$ defined in Lemma \ref{lemanalyticity} and corresponding to the second negative eigenvalue of $\Lg$ can be extended to $(-\infty,+\infty)$. Furthermore, $f(\g)\in\hu$ is an odd function, for each $\g\in(-\infty,+\infty)$.
\end{lem}

The proof uses a similar idea to that of Lemma \ref{lem:propnegev},
see Appendix~\ref{ap:impaire}.

We can deduce the number of negative eigenvalues of $\Lg$ in the radial case from the result on the eigenvalues of $\Lg$ considered in the whole space $\ld$. Indeed, Lemma \ref{lem:impaire} ensures that the second eigenvalue of $\Lg$ considered in the whole space $\ld$ is associated with an odd eigenvector, and thus disappears when the problem is restricted to subspace of radial functions. Furthermore,
since $\fg\in\hurad$ and $\dual{\Lg \fg}{\fg}<0$,
we can infer that the first negative eigenvalue of
$\Lg$ is still present when the problem is restricted to sets of radial functions. Recalling that $n(\Hg)=n(\Lg)$, we obtain.

\begin{lem}\label{lem:radialcase}
Let $\g<0$. Then the operator $\Hg$ considered on $\hurad$ has only one negative eigenvalue, that is $n(\Hg)=1$.
\end{lem}

Combining Lemma \ref{lem:radialcase}, Lemma \ref{lem:signdwgneg} and Proposition \ref{prop:criterion}, we recover the results of \cite{fj} recalled in Proposition \ref{prop:fj}.

Alternatively, subsection \ref{subsecpropnHwg} can be adapted to the radial case. All the function spaces should be reduced to spaces of even functions, and Lemma \ref{lem:radialcase} can also be proved in this way.

\section{Strong instability}\label{secblowup}

This section is devoted to the proof of Theorem \ref{thmblowup}. We
use the virial theorem (Proposition \ref{thmvirial}) whose
verification will be given in Appendix~\ref{secvirial}.

We begin by introducing some notations
$$
\M=\{ v\in\hurad\setminus\{0\};\K(v)=0,\Iwg(v)\leq 0 \},
$$
$$
\dM=\inf\{ \Swg(v);v\in \M \},
$$
where $\Swg$ and $\Iwg$ are defined in Proposition
\ref{propexistence} and $\K$ in Proposition \ref{thmvirial}.

Our proof is divided in three steps.

\bigskip

{\sc Step 1.} We prove that $\fwg$ is also a minimizer of $\dM$.

Because of Pohozaev identity $\K(\fwg)=0$ (see \cite{bl}), it is
clear that  $\dM\leq d(\w)$, thus we only have to show $\dM\geq
d(\w)$. Let $v\in \M$. If $\Iwg(v)=0$, we have $\Swg(v)\geq d(\w)$,
therefore we suppose $\Iwg(v)<0$. For $\a>0$, let $\va$ be such
that $\va(x)=\a^{1/2}v(\a x)$. We have
$$
\Iwg(\va)=\a^2\nldd{\dx v}+\w\nldd{v}-\g\a|v(0)|^2-\a^{(p-1)/2}\nlpsps{v},
$$
thus $\displaystyle\lim_{\a\goesto 0}\Iwg(\va)=\w\nldd{v}>0,$ and
by continuity there exists $0<\ao<1$ such that $\Iwg(\vao)=0$.
Therefore
\begin{equation}\label{etoile}
\Swg(\vao)\geq d(\w).
\end{equation}
Consider now $\displaystyle\frac{\partial}{\partial\a}\Swg(\va)=\a\nldd{\dx v}-\frac{\g}{2}|v(0)|^2-\frac{p-1}{2(p+1)}\a^{(p-3)/2}\nlpsps{v}.$
Since $p\geq 5$ and $\K(v)=0$, we have for $\a\in[0,1]$
$$
\frac{\partial}{\partial\a}\Swg(\va)\geq \a \K(v)-\frac{\g}{2}(1-\a)|v(0)|^2=-\frac{\g}{2}(1-\a)|v(0)|^2
$$
and thus $ \displaystyle\frac{\partial}{\partial\a}\Swg(\va)\geq
0\mbox{ for all }\a\in[0,1], $ which leads to $\Swg(v)\geq
\Swg(\vao)$. It follows by (\ref{etoile}) that $\Swg(v)\geq d(\w)$,
which concludes $\dM=d(\w)$.

\bigskip

{\sc Step 2.} We construct a sequence of initial data $\fwga$ satisfying the following properties :
$$
\Swg(\fwga)<d(\w), \Iwg(\fwga)<0\mbox{ and }\K(\fwga)<0.
$$
These properties are invariant under the flow of (\ref{nls}).

For $\a>0$, we define $\fwga$ by $\fwga(x)=\a^{1/2}\fwg(\a x)$.
Since $p\geq 5$, $\g<0$ and $\K(\fwg)=0$, easy computations permit to obtain
$$
\displaystyle\frac{\partial^2}{\partial\a^2}\Swg(\fwga)_{|\a=1}<0, \displaystyle\frac{\partial}{\partial\a}\Iwg(\fwga)_{|\a=1}<0 \mbox{ and }
\displaystyle\frac{\partial}{\partial\a}\K(\fwga)_{|\a=1}<0,
$$
and thus for any $\a>1$ close enough to $1$ we have
\begin{equation}\label{prop}
\Swg(\fwga)<\Swg(\fwg), \Iwg(\fwga)<0\mbox{ and }\K(\fwga)<0.
\end{equation}

Now fix a $\a>1$ such that (\ref{prop}) is satisfied, and let
$\ua(t,x)$ be the solution of (\ref{nls}) with $\ua(0)=\fwga$.
Since $\fwga$ is radial, $\ua(t)$ is also radial for all $t>0$ (see
Remark \ref{rmqradial}). We claim that the properties described in
(\ref{prop}) are invariant under the flow of (\ref{nls}). Indeed,
since from (\ref{conservationenergy}) and
(\ref{conservationcharge}) we have for all $t>0$
\begin{equation}\label{prop2}
\Swg(\ua(t))=\Swg(\fwga)<\Swg(\fwg),
\end{equation}
we infer that  $\Iwg(\ua(t))\neq 0$ for any $t\geq 0$, and by continuity  we have $\Iwg(\ua(t))< 0$ for all $t\geq 0$. It follows that $\K(\ua(t))\neq 0$ for any $t\geq0$ (if not $\ua(t)\in \M$ and thus $\Swg(\ua(t))>\Swg(\fwg)$
 which contradicts (\ref{prop2})), and by continuity we have $\K(\ua(t))<0$ for all $t\geq 0$.

\bigskip

{\sc Step 3.} We prove that $\K(\ua)$ stays negative and away from
$0$ for all $t\geq 0$.

Let $t>0$ be arbitrary chosen, define $v=\ua(t)$ and  for $\b>0$ let $\vb$ be such that $\vb(x)=v(\b x)$.
Then we have
$$
\K(\vb)=\b\nldd{\dx v}-\frac{\g}{2}|v(0)|^2-\b^{-1}\frac{p-1}{2(p+1)}\nlpsps{v},
$$
thus $\lim_{\b\goesto +\infty}\K(\vb)=+\infty$, and by continuity there exists $\bo$ such that $\K(\vbo)=0$.
If $\Iwg(\vbo)\leq 0$, we keep $\bo$ unchanged; otherwise, we replace it by $\bot$ such that $1<\bot<\bo$, $\Iwg(\vbot)=0$ and $\K(\vbot)\leq0$. Thus in any case we have $\Swg(\vbo)\geq d(\w)$. Now, we have
$$
\Swg(v)-\Swg(\vbo)=\frac{1-\bo}{2}\nldd{\dx v}+(1-{\bo}^{-1})\left( \frac{\w}{2} \nldd{v}-\frac{1}{p+1}\nlpsps{v} \right),
$$
from the expression of $\K$ and $\bo>1$ it follows that
\begin{equation}\label{prop3}
\Swg(v)-\Swg(\vbo)\geq \frac{1}{2}(\K(v)-\K(\vbo)).
\end{equation}
Therefore, from (\ref{prop3}), $\K(\vbo)\leq0$ and $\Swg(\vbo)\geq d(\w)$ we have
\begin{equation}\label{viriel}
\K(v)\leq -m=2(\Swg(v)-d(\w))<0
\end{equation}
where $m$ is independent of $t$ since $\Swg$ is a conserved quantity.

\bigskip

{\sc Conclusion.}
Finally, thanks to (\ref{viriel}) and Proposition \ref{thmvirial}, we have
\begin{equation}\label{viriel2}
\nldd{x\ua(t)}\leq -mt^2+Ct+\nldd{x\fwga}.
\end{equation}
For $t$ large, the right member of (\ref{viriel2}) becomes negative, thus there exists $\Ta<+\infty$ such that
$$
\lim_{t\goesto\Ta}\nldd{\dx \ua(t)}=+\infty.
$$
Since it is clear that $\fwga\goesto\fwg$ in $\hu$ when $\a\goesto 1$, Theorem \ref{thmblowup} is proved.

\begin{rmq}\label{globalinstability}
It is not hard to see that the set
$$
\mathcal{I}=\{ v\in\hu;\Swg(v)<d(\w),\Iwg(v)>0 \}
$$
is invariant under the flow of (\ref{nls}), and that a solution
with initial data belonging to $\mathcal{I}$ is global. Thus using
the minimizing character of  $\fwg$  and performing an analysis in
the same way than in \cite{gss}, it is possible to find a family
of initial data in $\mathcal{I}$ approaching $\fwg$ in $\hu$ and
such that the associated solution of (\ref{nls}) exists globally
but escaped in finite time from a tubular neighborhood of
$\fwg$(see also \cite{f,g} for an illustration of this approach on
a related problem).
\end{rmq}

\section{Numerical results}\label{sec:numerics}
In this Section, we use numerical simulations to complement the
rigorous theory on stability and instability of the standing waves
of~(\ref{nls}). Our approach here is similar to the one
in~\cite{fsw}. In order to study stability under radial
perturbations, we use the initial condition
\begin{equation}
u_{0}(x) = (1 + \delta_{p}) \fwg(x). \label{eq:initial_condition_1}
\end{equation}
In order to study stability under non-radial
(asymmetric) perturbations, we use the initial condition
\begin{equation}
u_{0}(x)= \fwg(x-\delta_{c}),
\label{eq:initial_condition_2}
\end{equation}
when $\delta_{c}$ is the lateral shift of the initial condition.
Since the evolution of the momentum for solutions of
Eq.~(\ref{nls-lin-pot}) is given by
\begin{eqnarray}
\frac{d \mathcal{M}}{dt} = - 2 \int |u|^2 \nabla V(x) dx,
\end{eqnarray}
one can see that symmetry-breaking
perturbations~(\ref{eq:initial_condition_2}) do not conserve the
momentum and thus, may give rise to drift instabilities. In some
cases (when the standing wave has a negative slope and the
linearized problem has two negative eigenvalues), we use the initial
condition
\begin{equation}
\label{eq:ic3} u_{0}(x) = (1 + \delta_p) \fwg(x-\delta_{c}).
\end{equation}
In order to demonstrate the agreement of the numerics with the
rigorous stability theory, one needs to observe that $\| u -
\fwg\|_{H^1}$ remains ``small'' in the case of stability but
increases in the case of instability. In the latter case, however,
observing numerically that $\| u - \fwg\|_{H^1}$ increases does not
enable us to distinguish between the different types of
instabilities such as total diffraction (i.e., when $\lim_{t \to
\infty} \| u \|_\infty = 0$), finite-width instability, strong
instability or drift instability. Therefore, instead of presenting
the $H^1$ norm, we plot the dynamics of the maximal amplitude of the
solution and of the location of the maximal amplitude. Together,
these two quantities give a more informative description of the
dynamics, while also showing whether the soliton is stable.

\subsection{Stability in $\hurad$}
\label{sec:radial_stability}

\subsubsection{Strength of radial stability}
\label{subsec:G_positive}

When $\gamma > 0$, the standing waves are known to be stable
in~$\hurad$ for $1 < p \le 5$. The rigorous theory, however, does
not address the issue of the {\em strength of radial stability}.
This issue is of most interest in the case $p = 5$, which is
unstable when $\gamma = 0$.

\begin{figure}[hp]
 \centerline{\scalebox{0.6}{\includegraphics{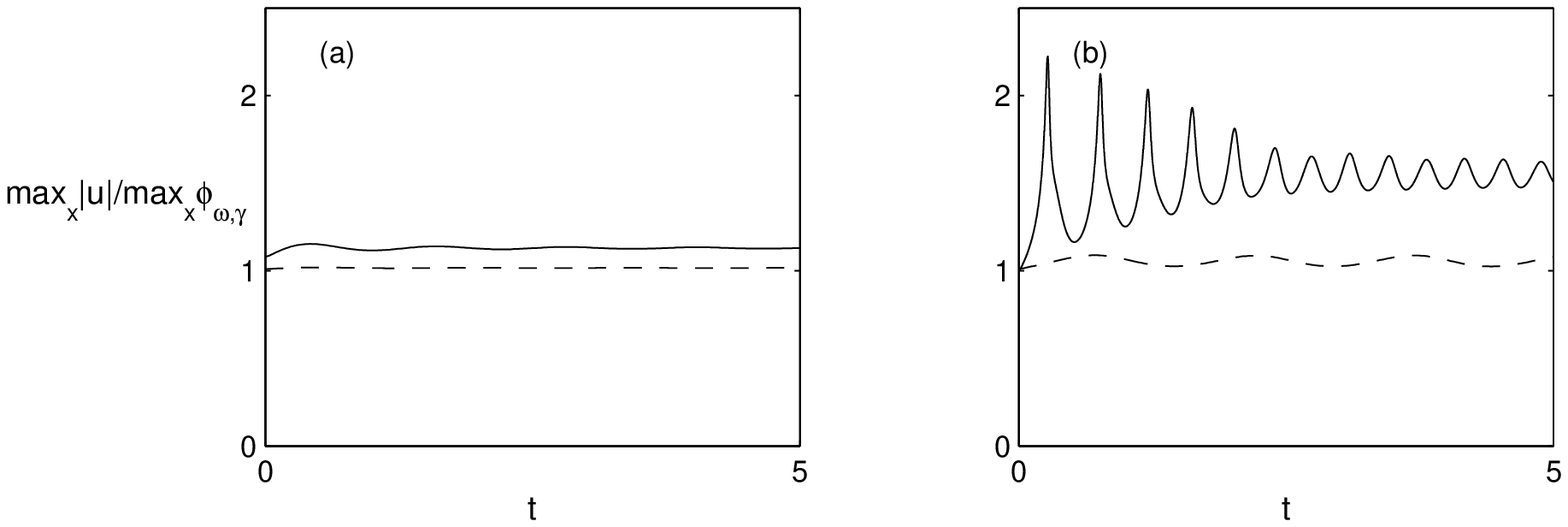}}}
  \caption{$max_{x}|u|/max_{x} \fwg$ as a function of $t$ for $\omega=4, \gamma=1$, $\delta_{p}=0.01$ (dashed line) and $\delta_{p}=0.08$ (solid line). (a) $p=3$ (b) $p=5$.}
    \label{fig:plot0102}
\end{figure}

For $\delta_{p}>0$, it is useful to define
\begin{equation}
F(\delta_{p})= \max_{t \ge 0} \left\{ {{\max_{x}|u(x,t)| -
\max_{x} \fwg} \over \max_{x}  \fwg} \right\}
\label{eq:omega_min_condition}
\end{equation}
as a measure of the strength of radial stability.
Figure~\ref{fig:plot0102} shows the normalized values ${\max_{x}
|u|}/{\max_{x}  \fwg}$ as a function of~$t$, for the initial
condition~(\ref{eq:initial_condition_1}) with
$\omega=4$ and $\gamma=1$. When $p=3$, a perturbation of
$\delta_{p}=0.01$ induces small oscillations and $F(0.01)=1.9\%$.
Therefore, roughly speaking, a 1\% perturbation of the initial condition leads to a
maximal deviation of 2\%.
A larger perturbation of $\delta_{p}=0.08$ causes the magnitude of the
oscillations to increase approximately by the same ratio, so that
$F(0.08)=15\%$. Using the same perturbations with $p=5$, however, leads to significantly
larger deviations. Thus,
$F(0.01)=8.8\%$, i.e., more than 4 times bigger than for $p=3$, and
$F(0.08)=122\%$, i.e., more than 8 times than for $p=3$.

In \cite{fsw,fsw2}, Fibich, Sivan and Weinstein observed that the
strength of radial stability is related to the magnitude of slope
$\partial_{\omega} {|| \fwg||^{2}_{2}}$, so that the
larger $\partial_{\omega} {|| \fwg||^{2}_{2}}$, the "more
stable" the solution is. Indeed, numerically we found that when
$\omega=4$, $\partial_{\omega} {|| \fwg||^{2}_{2}}$ is
equal to $ 1.0$ for $p=3$ and $0.056$ for $p=5$.

\begin{figure}[hp]
 \centerline{\scalebox{0.6}{\includegraphics{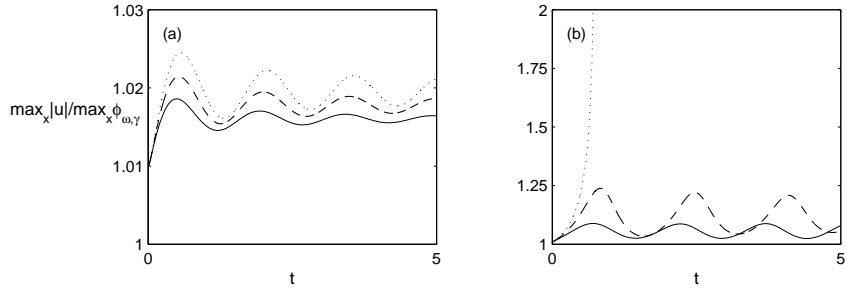}}}
  \caption{$max_{x}|u|/max_{x} \fwg$ as a function of $t$ for $\omega=4$,
$\delta_{p}=0.01$, and $\gamma=1$ (solid line), $\gamma=0.5$ (dashed line) and $\gamma=0.1$
(dots). (a) $p=3$ (b) $p=5$.}
    \label{fig:peak_as_function_gamma}
\end{figure}

Since when $\gamma = 0$, the slope
is positive for $p<5$ but zero for $p=5$, for $\gamma>0$
the slope is smaller in the critical case than in the subcritical case.
Therefore, we make the following informal observation:
\begin{observation}
    Radial stability of the standing waves of~$(\ref{nls})$ with
    $\gamma>0$ is ``weaker'' in the critical case $p=5$ than in the
subcritical case $p<5$.
\end{observation}
Clearly, this difference would be more dramatic at smaller (positive) values of~$\gamma$.
 Indeed, if in the simulation of Figure~\ref{fig:plot0102} with
$\delta_p = 0.01$ we reduce $\gamma$ from 1 to 0.5 and then to 0.1, this has almost no effect when $p=3$, where the value of~$F$ slightly increases from $1.9\%$ to $2.1\%$ and to $2.5\%$, respectively, see Figure~\ref{fig:peak_as_function_gamma}a. However,
if we repeat the same simulations with $p=5$, then reducing the value of $\gamma$
has a much larger effect, see Figure~\ref{fig:peak_as_function_gamma}b, where
$F$~increases from $8.9\%$ for $\gamma = 1$ to $24\%$
for $\gamma = 0.5$. Moreover, when we further reduced~$\gamma$  to~0.1, the solution seems to undergo collapse.\footnote{Clearly, one cannot use
numerics to determine that a solution becomes singular, as it is
always possible that collapse would be arrested at some higher
focusing levels.}
This implies that when $p=5$ and $\gamma>0$,
the standing wave is stable, yet it can collapse under a
sufficiently large perturbation.

\subsubsection{Characterization of radial instability for $3<p<5$ and $\gamma<0$}\label{subsec:G_negetive}

We consider the subcritical repulsive case  $p=4$ and $\gamma=-1$.  In this case,
there is threshold $\omega_{2}$ such that $\fwg$ is
stable for $\omega>\omega_{2}$ and unstable for
$\omega<\omega_{2}$. By numerical calculation we found that
$\omega_{2} (p=4, \gamma=-1) \approx 0.82$. Accordingly,
we chose two representative values of $\omega$: $\omega=0.5$ in the
unstable regime, and  $\omega=2$ in the stable regime.

\begin{figure}[hp]
    \centerline{\scalebox{0.5}{\includegraphics{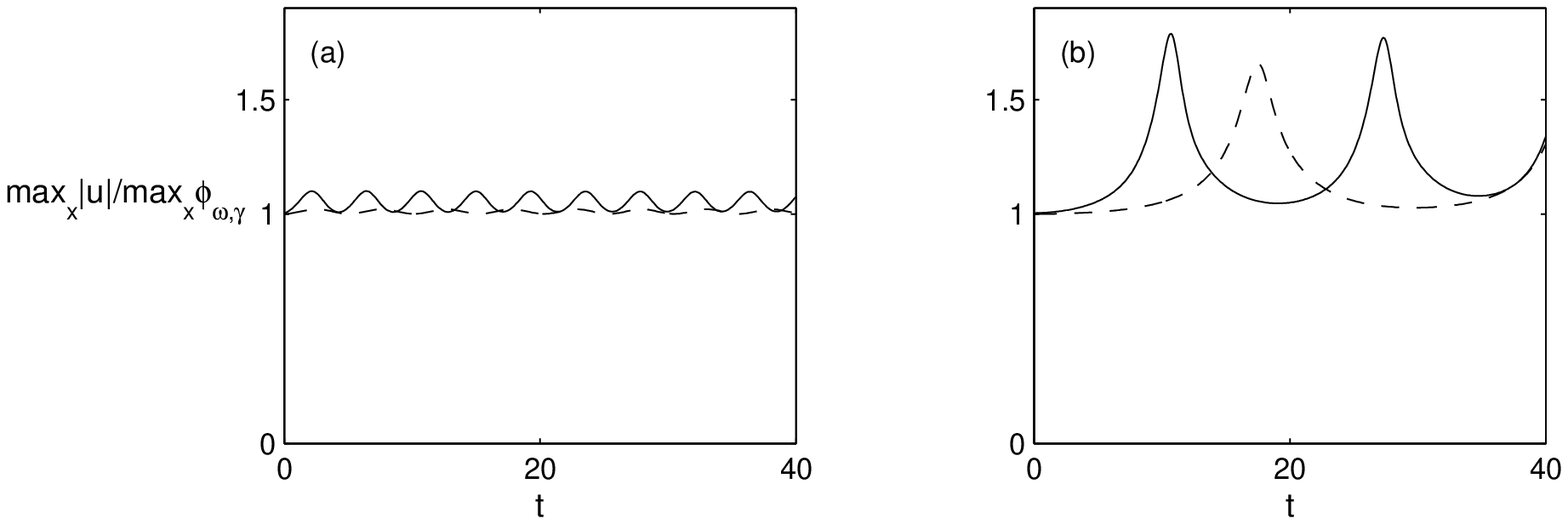}}}
  \caption{$max_{x}|u|/max_{x} \fwg$ as a function of $t$ for $p=4, \gamma=-1$, $\delta_{p}=0.001$ (dashed line) and $\delta_{p}=0.005$ (solid line). (a) $\omega=2$; (b) $\omega=0.5$.}
    \label{fig:plot0304}
\end{figure}

Figure~\ref{fig:plot0304}a demonstrates the stability
for~$\omega=2$. Indeed, reducing the perturbation from
$\delta_{p}=0.005$ to 0.001 results in reduction of the relative
magnitude of the oscillations by roughly five times, from
$F(0.005) \approx 10\%$ to $F(0.001) \approx 2\%$. The dynamics
in the unstable case $\omega=0.5$ is also oscillatory, see
Figure~\ref{fig:plot0304}b. However, in this case $F(0.005)=79\%$,
i.e., eight times larger than for~$\omega=2$. More importantly,
unlike the stable case, a perturbation of $\delta_{p}=0.001$ does
not result in a reduction of the relative magnitude of the
oscillations by $\approx 5$. In fact, the relative magnitude of
the oscillations decreases only to $F(0.001)=66\%$.

\begin{figure}[hp]
 \centerline{\scalebox{0.6}{\includegraphics{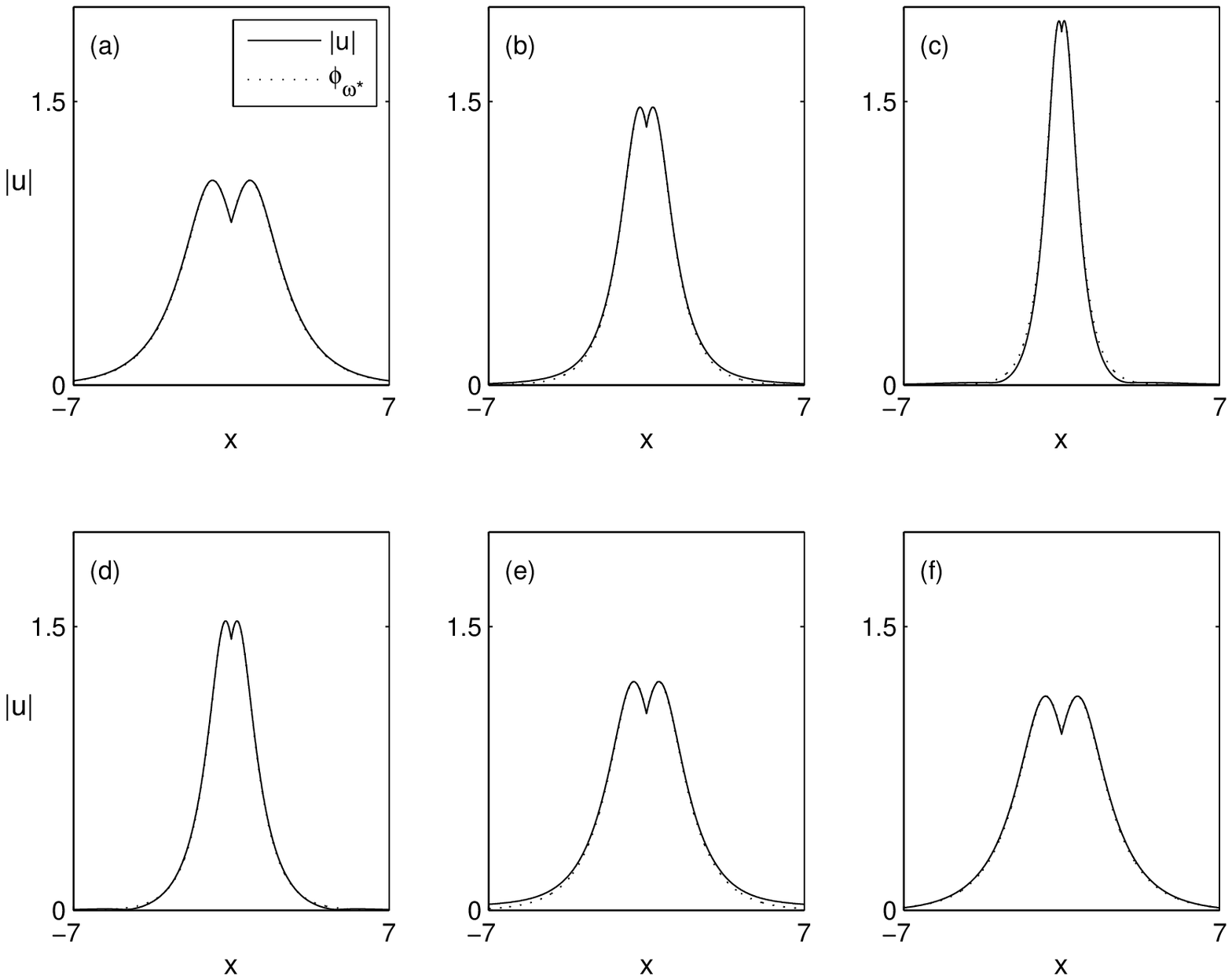}}}
  \caption{$|u(x,t)|$ (solid line) and $\phi_{\omega*(t)}(x)$ (dots) as a function of~$x$ for the simulation of Fig.~\ref{fig:plot0304}b with $\delta_{p}=0.005$. (a) $t=0~(\omega^{*}=0.508)$ (b) $t=9~(\omega^{*}=1.27)$ (c) $t=10.69~(\omega^{*}=2.86)$ (d) $t=12~(\omega^{*}=1.43)$ (e) $t=15~(\omega^{*}=0.706)$ (f) $t=20~(\omega^{*}=0.58)$.}
    \label{fig:plot04_evol}
\end{figure}

In the homogeneous NLS, unstable standing waves perturbed with $\delta_p>0$
always undergo collapse. Since, however, for $p=4$ it is impossible to have
collapse, an interesting question is the nature of the instability
in the unstable region $\omega<\omega_{2}$. In Figure~\ref{fig:plot0304}b we already saw
that $max|u(x,t)|$~undergoes oscillations. In order
to better understand the nature of this unstable oscillatory dynamics, we plot
in Figure~\ref{fig:plot04_evol} the spatial profile of~$|u(x,t)|$
at various values of~$t$. In addition, at each~$t$ we plot
$\phi_{{\omega}^{*}(t), \gamma}(x)$, where $\omega^{*}(t)$ is determined
from the relation
\begin{eqnarray*}
    \max_{x} \phi_{\omega^{*}(t),\gamma}(x) = \max_{x} |u(x,t)|.
\end{eqnarray*}
Since the two curves are nearly indistinguishable (especially in the central
region), this shows that the unstable dynamics corresponds to
"movement along the curve $\phi_{\omega*(t)}$".

\begin{figure}[hp]
 \centerline{\scalebox{0.6}{\includegraphics{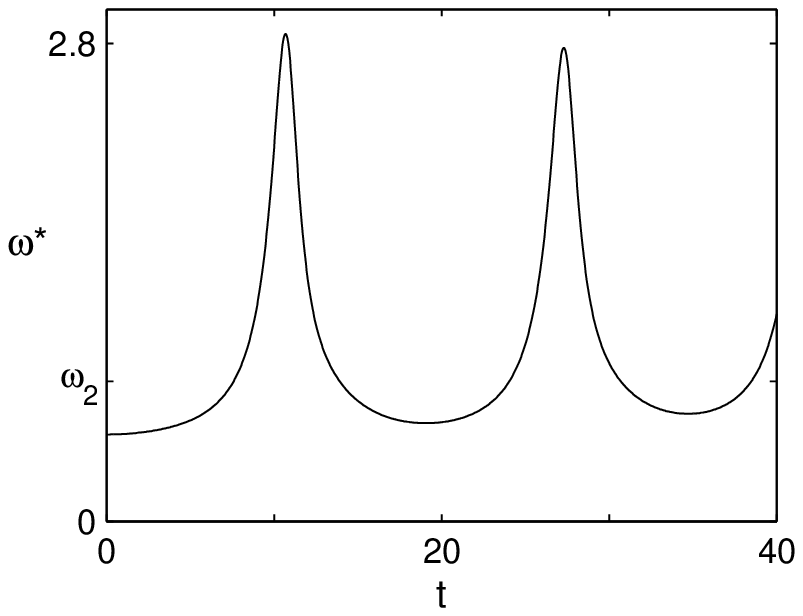}}}
  \caption{$\omega^{*}$ as  a function of $t$ for the simulation of Fig~\ref{fig:plot04_evol}.}
    \label{fig:omega_function2}
\end{figure}

In Figure~\ref{fig:omega_function2} we see that $\omega^{*}(t)$ undergoes oscillations, in accordance
with the oscillations of $\max_{x}|u|$. Furthermore, as one may
expect, collapse is arrested only when $\omega^{*}(t)$ reaches a
value $(\approx 2.86)$ which is in the stability region (i.e.,
above $\omega_{2}$).

\begin{observation}
 When $\gamma<0$ and $3<p<5$, the instability in $\hurad$ is a "finite width
instability", i.e., the solution narrows down along the curve
$\phi_{\omega*(t), \gamma}$ until it "reaches" a finite width in the stable region
$\omega>\omega_{2}$, at which point collapse is arrested.
\end{observation}
Note that this behavior was already observed in \cite{fsw}, Fig~19. Therefore, more generally, we conjecture that
\begin{observation}
When the slope is negative
$($i.e., $\partial_{\omega} {|| \fwg||^{2}_{2}}<0$ $)$,
then the symmetric perturbation~$(\ref{eq:initial_condition_1})$
with $0<\delta_p \ll 1$
leads to a finite-width instability in the subcritical case, and to a
finite-time collapse in the critical and supercritical cases.
\end{observation}

\subsubsection{Supercritical case $(p>5)$}\label{Super_Critical_Case}

\begin{figure}[hp]
 \centerline{\scalebox{0.6}{\includegraphics{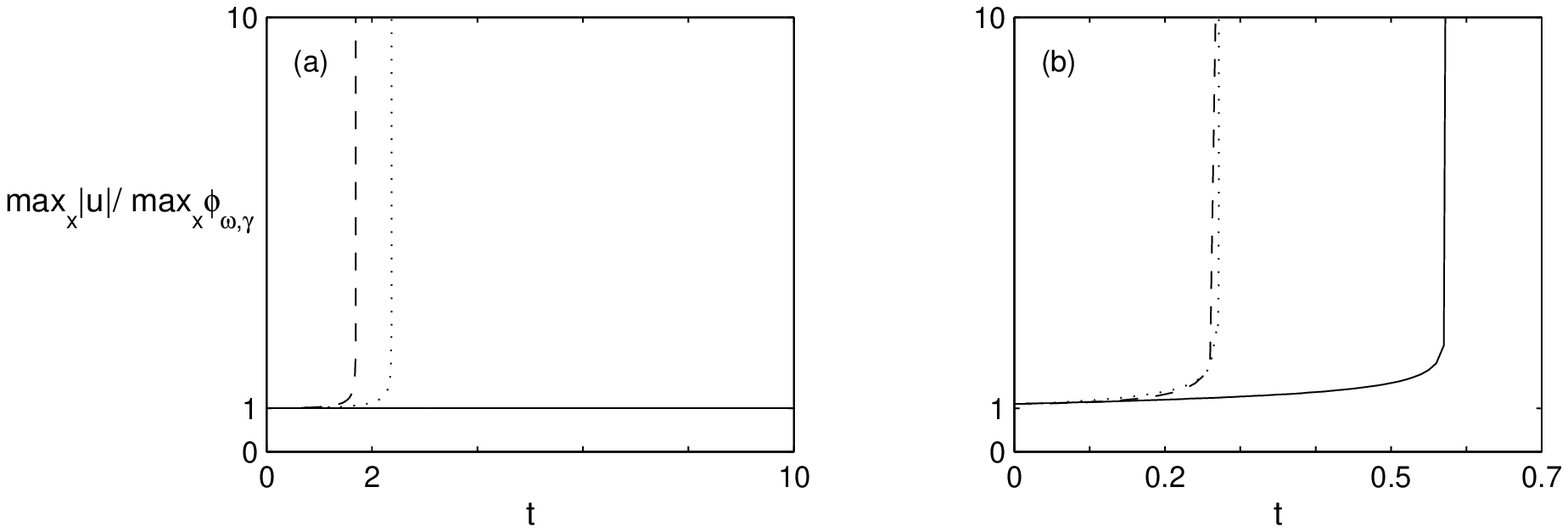}}}
  \caption{$\max_{x}|u(x,t)|/ \max_{x} \fwg$ as a function of $t$ for $p=6, \omega=1$ and $\gamma=-1$ (dashed line), $\gamma=0$ (dots), $\gamma=+1$ (solid line). (a) $\delta_{p}=0.001$ (b) $\delta_{p}=0.1$.}
    \label{fig:super_critical_case}
\end{figure}

\begin{figure}[hp]
 \centerline{\scalebox{0.6}{\includegraphics{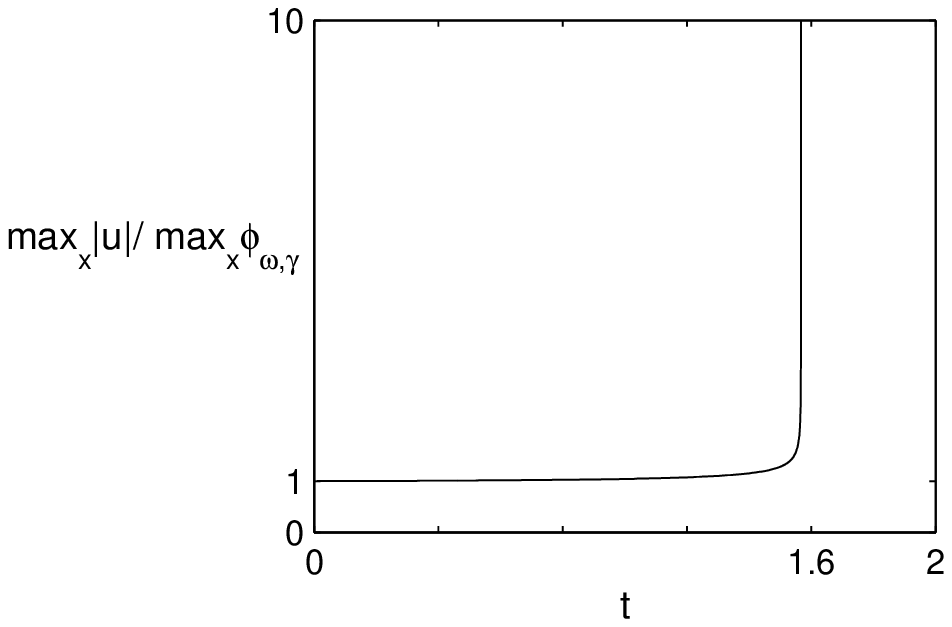}}}
  \caption{$\max_{x}|u(x,t)|/ \max_{x} \fwg$ as a function of $t$ for $p=6, \omega=4$, $\gamma=1$ and $\delta_{p}=0.001$.}
    \label{fig:super_critical_omega4}
\end{figure}

We recall that when $\gamma>0$ and $p>5$, the standing wave is stable
for $\gamma^{2}/4<\omega<\omega_{1}$ and unstable for
$\omega_{1}<\omega$. When $\gamma<0$ and $p>5$ the standing wave is strongly
unstable under radial perturbations for any $\omega$, i.e., an
infinitesimal perturbation can lead to collapse.
\\Figure~\ref{fig:super_critical_case} shows the behavior of
perturbed solutions for $p=6$ and $\omega=1$. As predicted by the
theory, when $\delta_{p}=0.001$, the solution blows up for
$\gamma=-1$ and $\gamma=0$, but undergoes small oscillations (i.e., is
stable) for $\gamma=1$. Indeed, we found numerically that
$\omega_{1}(p=6,\gamma=1) \approx 2.9$, so that the
standing wave is stable for $\omega=1$. However, when we increase the
perturbation to $\delta_{p}=0.1$, the solution with $\gamma=1$
also seems to undergo collapse. This implies that when $p>5, \gamma>0$ and
$\omega<\omega_1$ the standing wave is stable, yet it can collapse under a
sufficiently large perturbation. In order to find the type of
instability for $\gamma>0$ and $\omega>\omega_{1}$, we solve
the NLS~(\ref{nls}) with $p=6$, $\gamma=1$ and $\omega=4$. In this
case, $\delta_{p}=0.001$ seems to lead to collapse, see
Figure~\ref{fig:super_critical_omega4}, suggesting a strong
instability for  $p>5$, $\gamma>0$ and $\omega>\omega_{1}$.
Therefore, we make the following informal observation:
\begin{observation}
If a standing wave of~$(\ref{nls})$ with $p > 5$ is unstable in~$\hurad$, then
the instability is strong.
\end{observation}

\subsection{Stability under non-radial perturbations}\label{sec:devision}

\subsubsection{Stability for $1<p<5$ and $\gamma>0$}\label{subsec:G_positive_drift}

\begin{figure}[hp]
 \centerline{\scalebox{0.6}{\includegraphics{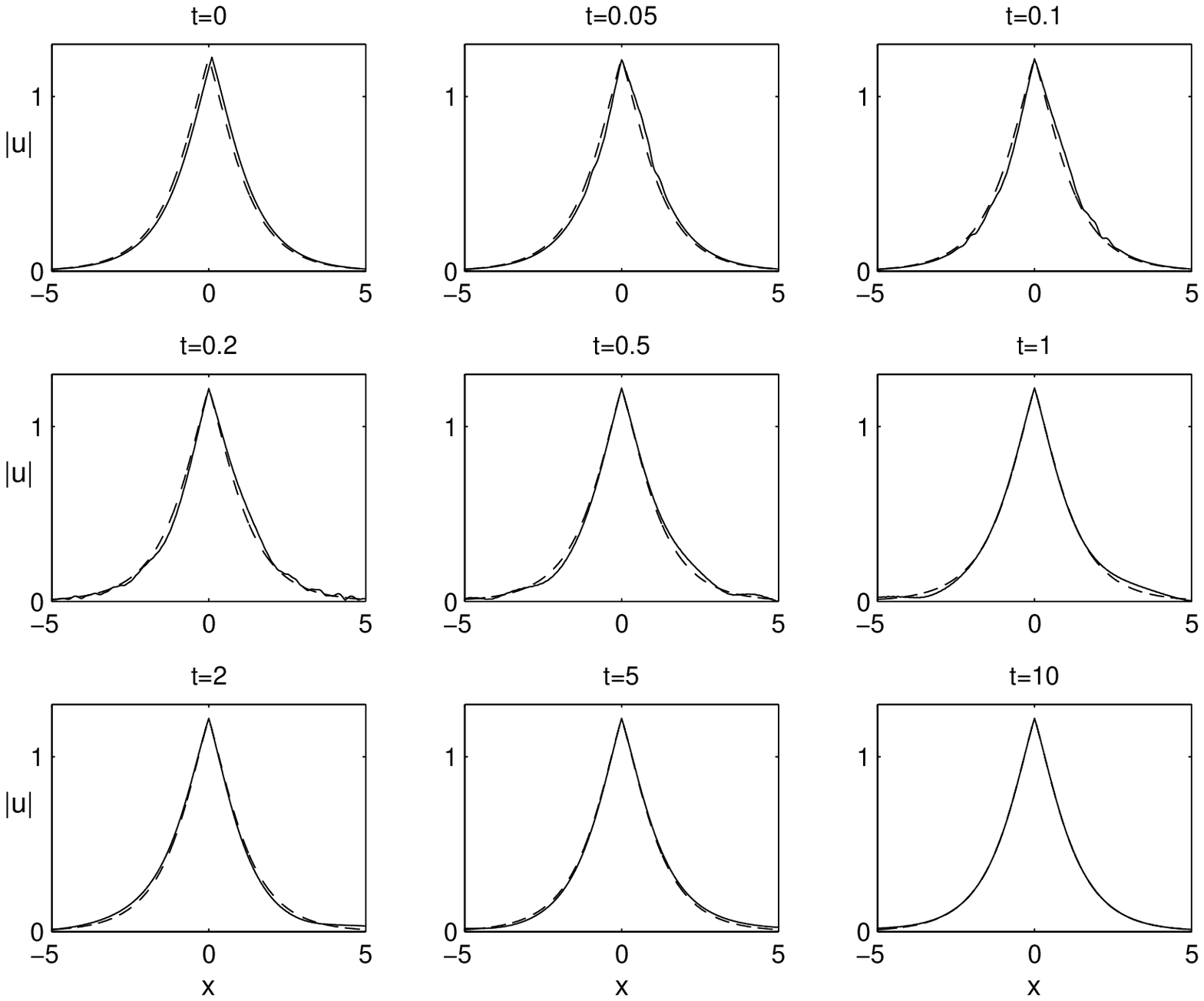}}}
  \caption{$|u(x,t)|$ (solid line) and $\phi_{\omega*=0.995}(x)$ (dashed line) as a function of~$x$. Here, $p=3, \omega=1, \gamma=1$ and $\delta_{c}=0.1$.}
    \label{fig:stable_drift_evolution}
\end{figure}

\begin{figure}[hp]
 \centerline{\scalebox{0.6}{\includegraphics{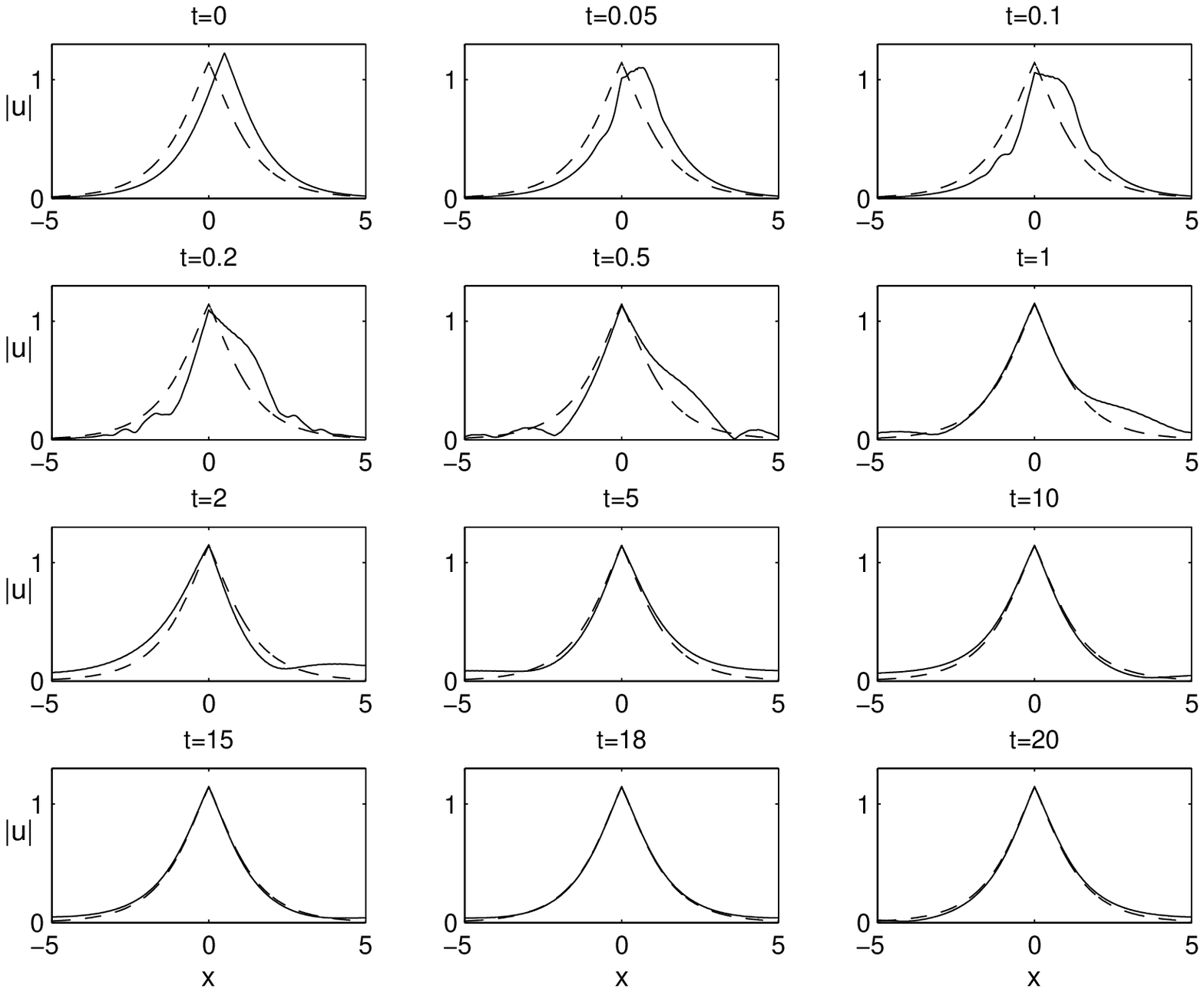}}}
  \caption{Same as Fig~\ref{fig:stable_drift_evolution} with $\delta_{c}=0.5$ and $\omega^*=0.905$.}
    \label{fig:stable_drift_evolution_Ec05}
\end{figure}

Figure~\ref{fig:stable_drift_evolution} shows the evolution of the
solution when $p=3,  \gamma=1, \omega=1$ and $\delta_{c}=0.1$. The
peak of the solution moves back towards $x=0$ very quickly (around
$t \approx 0.003$) and stays there at later times. Subsequently,
the solution converges to the bound state
$\phi_{\omega^{*}=0.995}$. This convergence starts near $x=0$ and
spreads sideways, accompanied by radiation of the excess power
$||u_{0}||_{2}^{2}-||\phi_{\omega^{*}=0.995}||_{2}^{2} \cong 2.00 -
1.99 = 0.01$. In Fig~\ref{fig:stable_drift_evolution_Ec05} we
repeat this simulation with a larger shift of $\delta_{c}=0.5$.
The overall dynamics is similar: The solution peak moves back to $x=0$,
and the solution converges (from the center outwards) to
$\phi_{\omega^{*}=0.905}$. In this case, it takes longer for the
maximum to return to $x=0$ (at $t \approx 0.11$), and more power is radiated
in the process ($||u_{0}||_{2}^{2}-||\phi_{\omega^{*}=0.905}||_{2}^{2} \cong 2.00 -
1.81 = 0.19$.  We verified that
the "non-smooth" profiles (e.g., at $t=0.2$) are not numerical
artifacts.

\subsubsection{Drift instability for $1<p \le 3$ and $\gamma<0$}
\label{subsec:G_negative_drift}

\begin{figure}[hp]
 \centerline{\scalebox{0.6}{\includegraphics{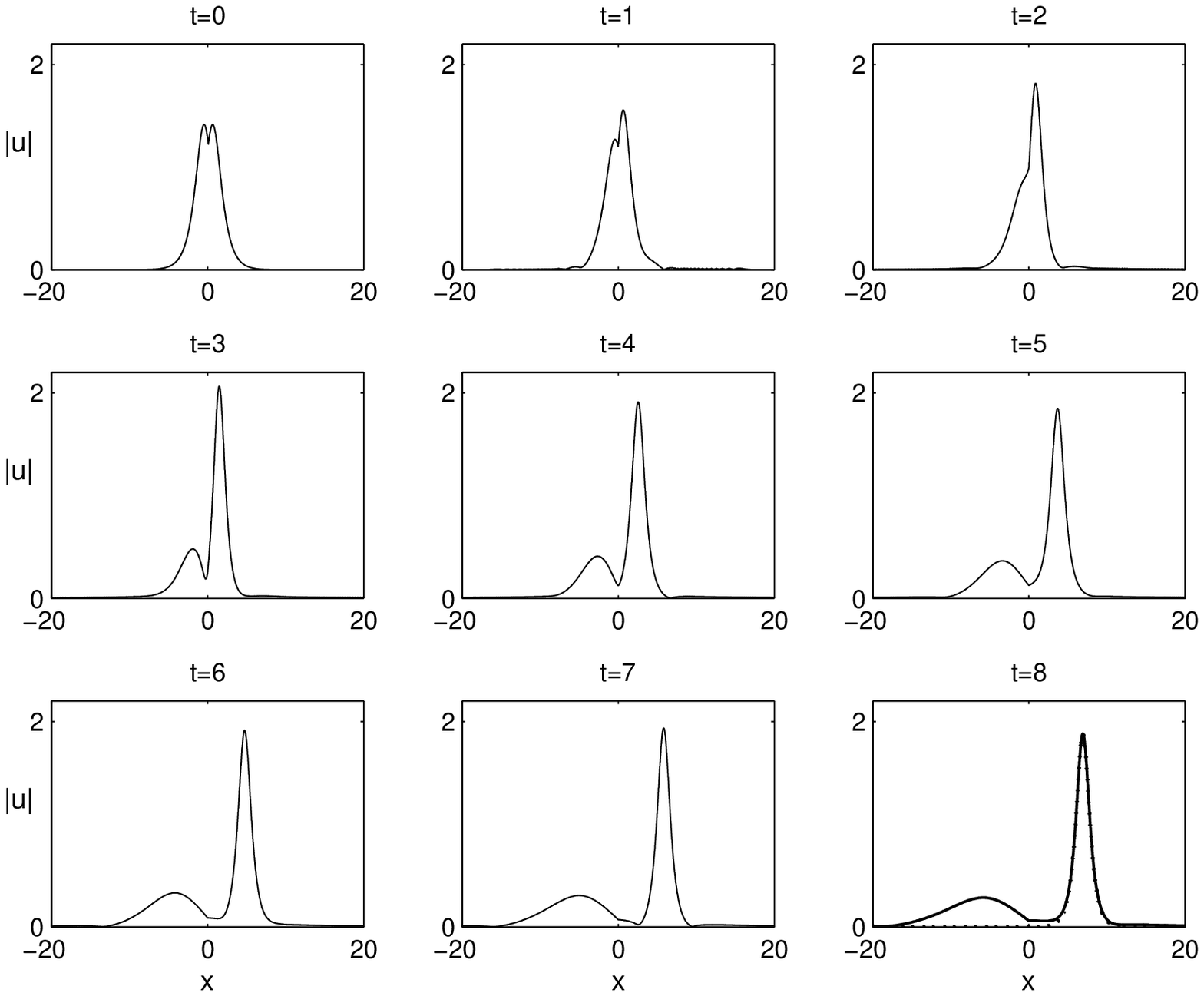}}}
  \caption{$|u(x,t)|$ (solid line) as a function of~$x$. Here $p=3, \gamma=-1, \omega=1$ and $\delta_{c}=0.1$. Dotted line at $t=8$ is $\sqrt{2\omega^{*}}sech(\sqrt{\omega^{*}}(x-x^*))$ with $\omega^{*}=1.768$ and
$x^{*} \approx 7$.}
    \label{fig:nonstable_drift_evolution_A}
\end{figure}

\begin{figure}[hp]
 \centerline{\scalebox{0.6}{\includegraphics{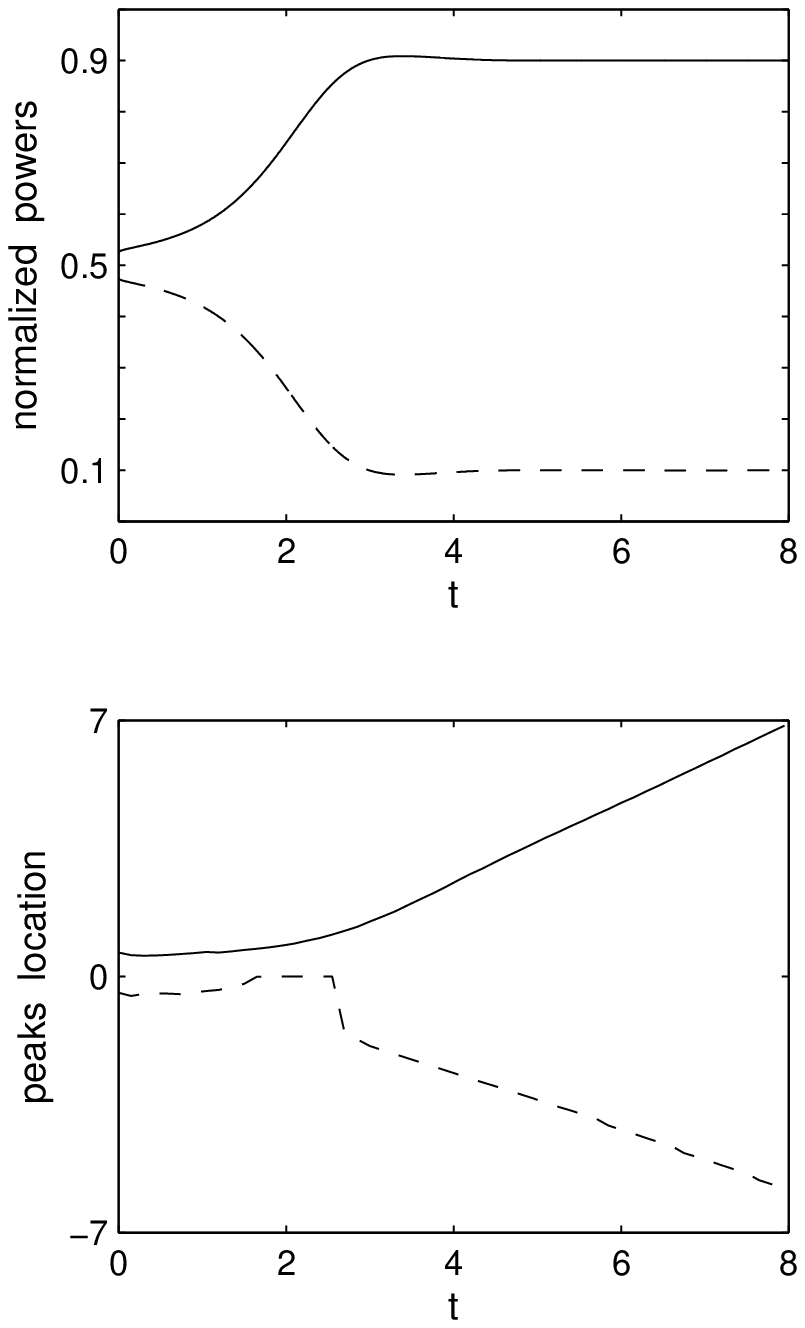}}}
  \caption{(a) The normalized powers $\int_{0}^{\infty}|u|^{2}dx / \int_{-\infty}^{\infty}|u_{0}|^{2}dx$ (solid line) and
$\int_{-\infty}^{0}|u|^{2}dx /
\int_{-\infty}^{\infty}|u_{0}|^{2}dx$ (dashed line), and (b)
location of $\max_{0 \le x}|u(x,t)|$ (solid line)
and of $\max_{x \le 0}|u(x,t)|$ (dashed line),
for the simulation of
Figure~\ref{fig:nonstable_drift_evolution_A}.}
    \label{fig:nonstable_drift_Power_flow_and_peak_location}
\end{figure}

Figure~\ref{fig:nonstable_drift_evolution_A} shows the evolution of the solution
for $p=3, \gamma=-1,  \omega=1$ and $\delta_{c}=0.1$.
Unlike the attractive case with the same parameters (Figure~\ref{fig:stable_drift_evolution}),
as a result of this small initial shift to the right, nearly all the power flows from the left side
of the defect $(x<0)$ to the right side $(x>0)$, see
Figure~\ref{fig:nonstable_drift_Power_flow_and_peak_location}a, so
that by $t \approx 3$, $\approx90\%$ of the power is in the right side.
Subsequently, the right component moves to the right at a constant
speed (see
Fig~\ref{fig:nonstable_drift_Power_flow_and_peak_location}b) while
assuming the $sech$ profile of the homogeneous NLS bound state
(see Fig~\ref{fig:nonstable_drift_evolution_A} at t=8);
 the left component also drifts away from the defect.

We thus see that
\begin{observation}
   When $1 <  p \le 3$, the standing waves are stable under shifts in the attractive case, but
undergo a drift instability away from the defect in the repulsive case.
\end{observation}
 We note that a similar behavior was observed in the subcritical NLS with a periodic nonlinearity,
see~\cite{fsw}, Section 5.1.

\subsubsection{Drift and finite-width instability for $3<p <5$ and $\gamma<0$}

\begin{figure}[hp]
 \centerline{\scalebox{0.6}{\includegraphics{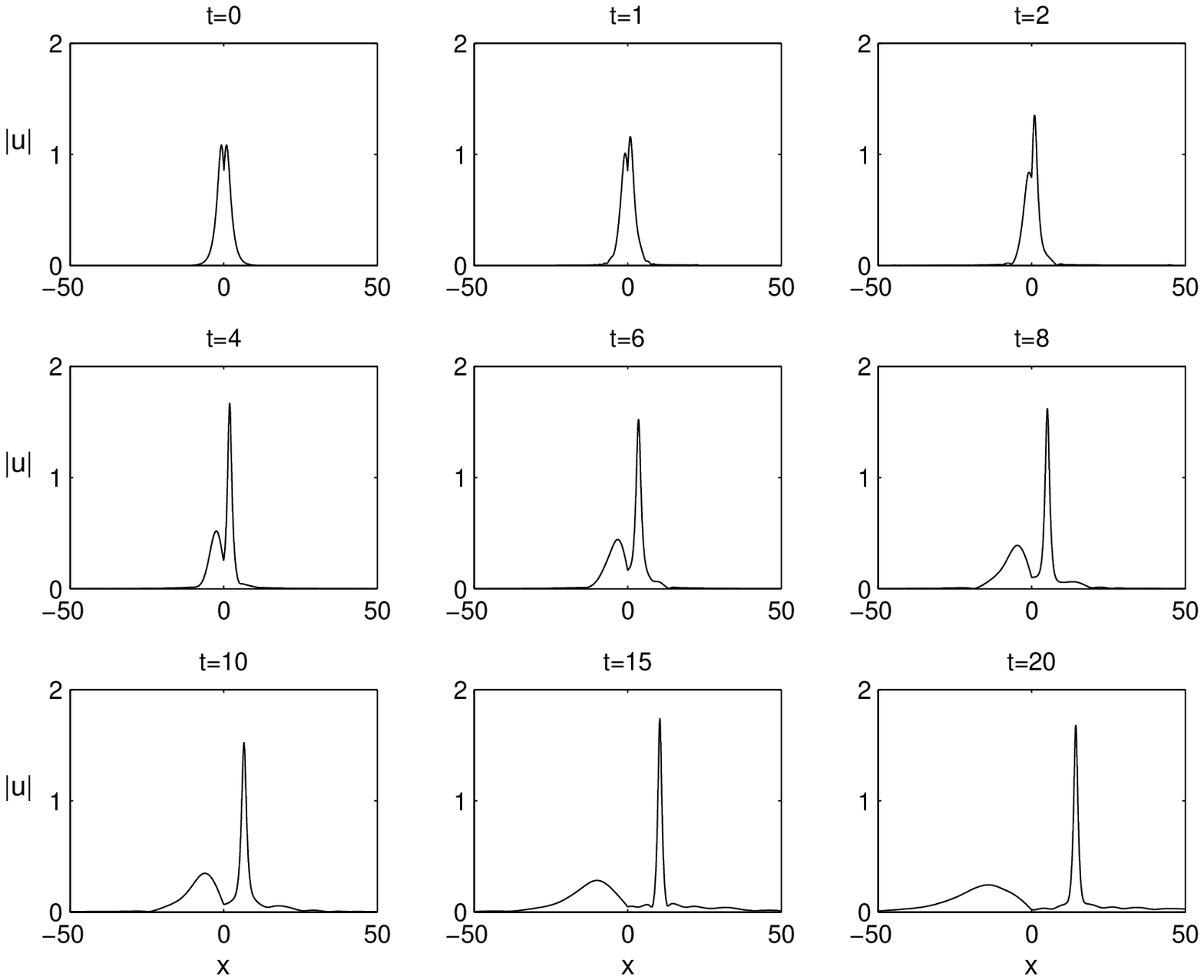}}}
  \caption{$u(x,t)$ as a function of $x$. Here $p=4$, $\gamma=-1$,  $\w=0.5$, $\delta_p=0.005$, and $\delta_c = 0.1$}
    \label{fig:plot_subcritical_drift_frames.eps}
\end{figure}

\begin{figure}[hp]
 \centerline{\scalebox{0.6}{\includegraphics{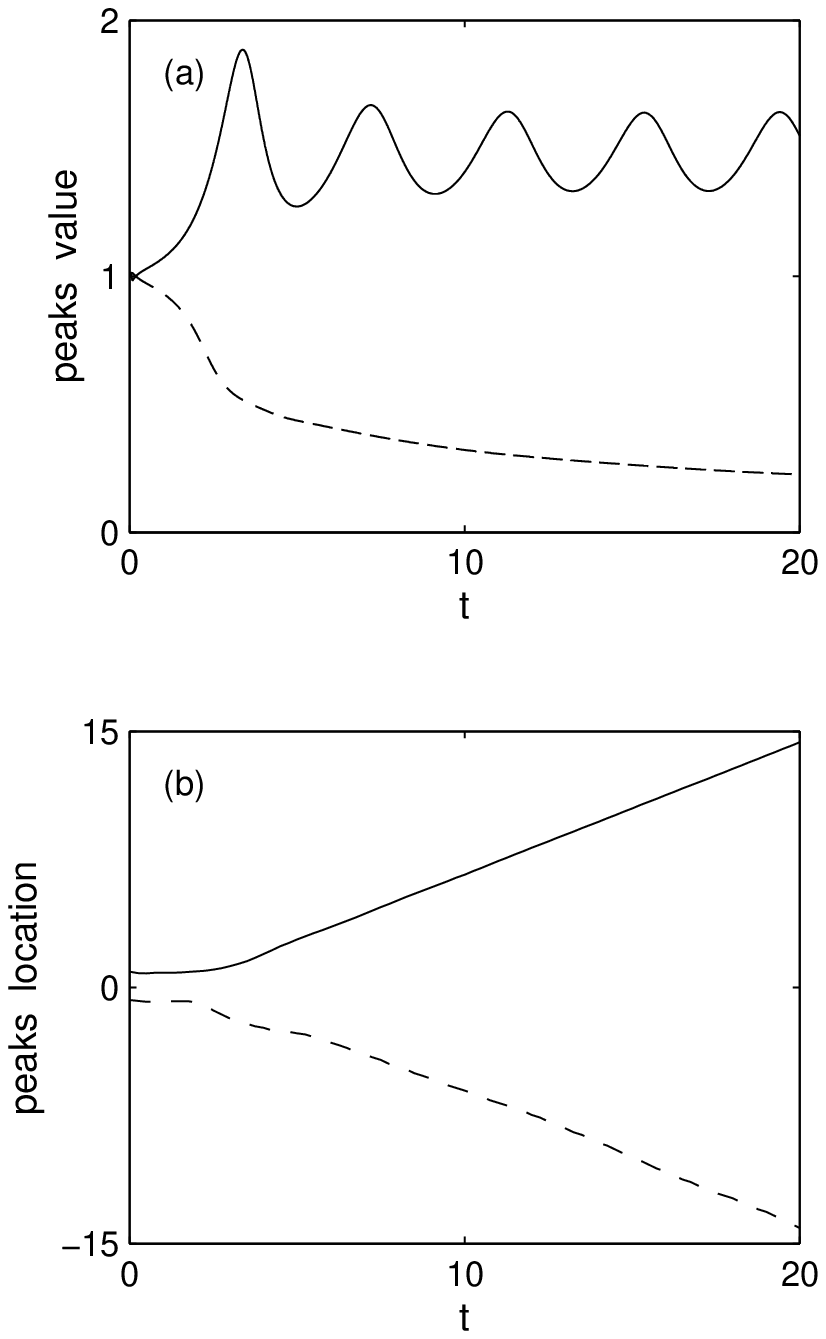}}}
  \caption{(a) The value, and (b) the location,
 of the right peak $\max_{0 \le x}|u(x,t)|$ (solid line)
and left peak  $\max_{x \le 0}|u(x,t)|$ (dashed line),
for the simulation of
Figure~\ref{fig:plot_subcritical_drift_frames.eps}.}
    \label{fig:plot_subcritical_drift_peak.eps}
\end{figure}

   In Figure~\ref{fig:plot0304}b, Figure~\ref{fig:plot04_evol}, and
Figure~\ref{fig:omega_function2}
 we saw that when $p=4$, $\gamma=-1$,  $\w=0.5$, and $\delta_p=0.005$,
the solution undergoes a finite-width instability in~$\hurad$. In
Figures~\ref{fig:plot_subcritical_drift_frames.eps}
and~\ref{fig:plot_subcritical_drift_peak.eps}  we show the
dynamics (in~$\hu$) when we add a small shift of $\delta_c = 0.1$.
In this case, the (larger) right component undergoes a combination
of a drift instability and a finite-width instability, whereas the
(smaller) left component undergoes a drift instability. Therefore,
we make the following observation
\begin{observation}
 When  $3<p <5$, $\gamma^2/4<\w < \w_2$ and $\gamma<0$, the standing waves
undergo a combined drift and finite-width instability.
\end{observation}

\subsubsection{Drift and strong instability for  $5 \le p$ and $\gamma<0$}

\begin{figure}[hp]
 \centerline{\scalebox{0.6}{\includegraphics{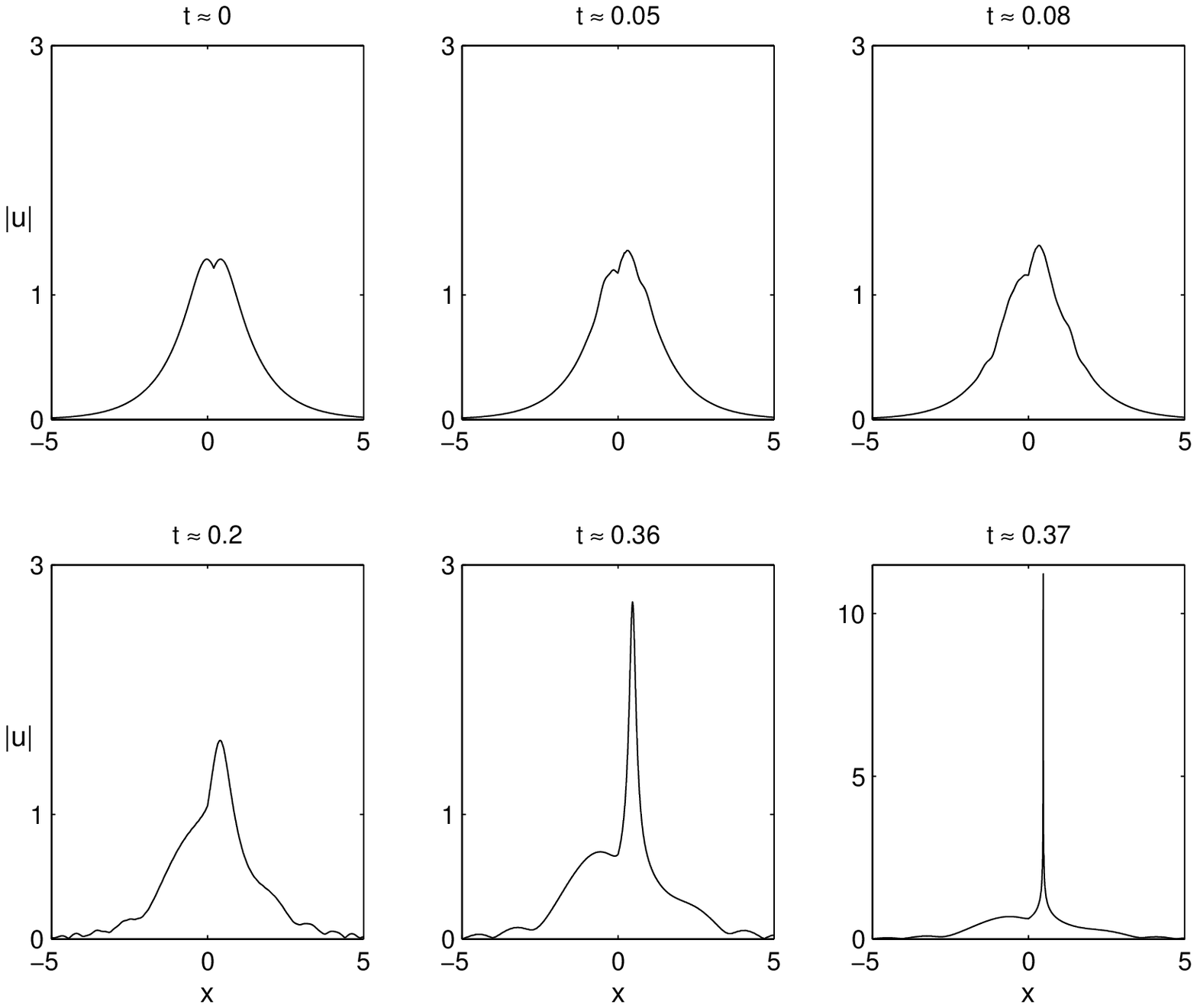}}}
  \caption{ $|u(x,t)|$ as a function of $x$, at various values of~$t$.
   Here, $p=6$, $\gamma=-1$, $\omega=1$, $\delta_{c}=0.2$ and $\delta_{p}=0.001$.}
    \label{fig:unstable_drift_and_peak_frames}
\end{figure}

\begin{figure}[hp]
 \centerline{\scalebox{0.7}{\includegraphics{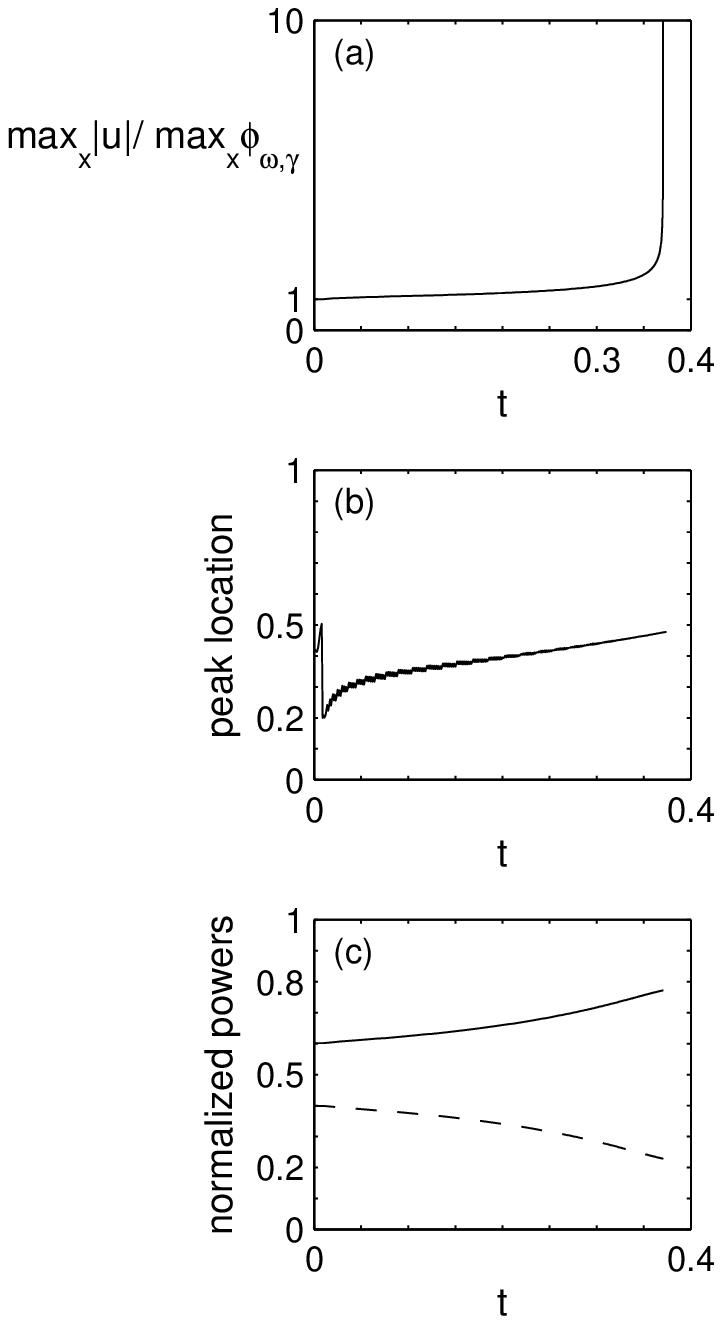}}}
  \caption{(a) $\max_{x}|u(x,t)|/ \max_{x}\fwg$ (b) location of $\max_{x}|u(x,t)|$ and (c)
The normalized powers $\int_{0}^{\infty}|u|^{2}dx / \int_{-\infty}^{\infty}|u_{0}|^{2}dx$ (solid line) and
$\int_{-\infty}^{0}|u|^{2}dx /
\int_{-\infty}^{\infty}|u_{0}|^{2}dx$ (dashed line),
for the solution of Fig.~\ref{fig:unstable_drift_and_peak_frames}.
}
    \label{fig:unstable_drift_and_peak_and_power}
\end{figure}

In Figures~\ref{fig:unstable_drift_and_peak_frames} and~\ref{fig:unstable_drift_and_peak_and_power}
we show the solution of the NLS~(\ref{nls}) with $p=6$, $\gamma=-1$ and $\omega=1$, for the initial
condition~(\ref{eq:ic3}) with $\delta_{c}=0.2$ and $\delta_{p}=0.001$.
As predicted by the theory, this strongly unstable solution undergoes collapse. Note, however,
that, in parallel, the solution also undergoes a drift instability.
We thus see that
\begin{observation}
   In the critical and supercritical repulsive case, the standing waves collapse
while undergoing a drift instability away from the defect.
\end{observation}
Note that a similar behavior was observed in \cite{fsw}, Section~5.2.

\subsection{Numerical Methods}
We solve the NLS~(\ref{nls}) using fourth-order finite differences
in~$x$ and  second-order implicit
Crack-Nicholson scheme in time. Clearly, the main question is how
to discretize the delta potential at $x=0$. Recall that in
continuous case
\begin{equation*}
\lim_{x\to0^{+}} \partial_{x} u(x) - \lim_{x\to0^{-}}
\partial_{x} u(x) = -\gamma u(0). \label{eq:discrete_term}
\end{equation*}
Discretizing this relation with $\mathcal{O}(h^2)$ accuracy gives
\begin{equation*}
{{u(2h)-4u(h)+3u(0)} \over 2h} - {{-u(-2h)+4u(-h)-3u(0)} \over 2h}
 = -\gamma u(0), \label{eq:discrete_term}
\end{equation*}
when $h$ is the spatial grid size. By rearrangement of the terms
we get the equation
\begin{equation}
-u(2h)+4u(h)+[2h \gamma-6]u(0)+4u(-h)-u(-2h)=0.
\label{eq:discrete_condition}
\end{equation}
When we simulate symmetric perturbations (section~\ref{sec:radial_stability}), we enforce
symmetry by solving only on half space $[0,+\infty)$. In this case, because
of the symmetry condition $u(-x)=u(x)$, (\ref{eq:discrete_condition}) becomes
$$
[2h\gamma-6]u(0)+8u(h)-2u(2h) = 0.
$$

\appendix \section{Proof of Lemma~\ref{lemmadefLug}}\label{ap:lemmadefLug}
The proof for $\Ldg$ being similar to the one of $\Lug$ we only deal
with $\Lug$. The form $\Bug$ can be decomposed into
$\Bug=\Buug+\Budg$ with $\Buug:\hu\times\hu\goesto\R$ and
$\Budg:\ld\times\ld\goesto\R$  defined by
\begin{equation}\label{eqBuugBudg}
\begin{array}{rcl}
\Buug(v,z)&=&\intr{\dx v\dx z dx}-\g v(0)z(0),\\
\Budg(v,z)&=&\w\intr{vz dx}-\intr p\fwg^{p-1}vzdx.
\end{array}
\end{equation}

If we denote by $\Tu$ (resp. $\Td$) the self-adjoint operator
on $\ld$ associated with $\Buug$ (resp $\Budg$),
it is clear that $D(\Td)=\ld$ and
$$
D(\Lug)=D(\Tu).
$$
If we take $v\in H^2(\R)$ such that $v(0)=0$,
and put $w=-\dxx v\in\ld$, it follows that for any $z\in\hu$
we have
$$
\Buug(v,z)=\intr{\dx v\dx z dx}=\psld{w}{z}.
$$
Thus $v\in D(\Tu)$, and
we can deduce that $\Tu$ is a self-adjoint extension
of the operator $T$ defined by
$$
T=-\dxx,\;~ D(T)=\{ v\in H^2(\R);~v(0)=0 \}.
$$
On the other hand,
using the theory of self-adjoint extensions of symmetric operators,
one can see (see \cite[Theorem I-3.1.1]{aghh})
that there exists $\a\in\R$ such that
$$
D(\Tu)=\{v\in\hu\cap H^2(\R\setminus\{0\});~\dx v(0^+)-\dx v(0^-)
=-\a v(0)\}.
$$
Now, take $v\in D(\Tu)$ with $v(0)\neq 0$. Then
\begin{eqnarray*}
\psld{\Tu v}{v}&=&\int^0_{-\infty}(-\dxx v)vdx+\int_0^{+\infty}(-\dxx v)vdx \\
&=&-v(0)\dx v(0-)+\int^0_{-\infty}|\dx v|^2dx+v(0)\dx v(0+)+\int_0^{+\infty}|\dx v|^2dx\\
&=&\intr|\dx v|^2dx-\a v(0)^2
\end{eqnarray*}
which should be equal to
$$
\Buug(v,v)=\intr|\dx v|^2dx-\g v(0)^2.
$$
Thus $\g=\a$, and the lemma is proved.

\section{Proofs of Lemmas \ref{lemma28} and
\ref{lemanalyticityLug}}\label{ap:28-analyticityLug}

{\it Proof of Lemma \ref{lemma28}.}
We start by showing that (i) and (iii) are satisfied. We work on
$\Lug$ and $\Ldg$. The essential spectrum of $\Tu$ (see the proof of
Lemma \ref{lemmadefLug}) is $\s_{\rm{ess}}(\Tu)=[0,+\infty)$. This
is standard when $\g=0$ and a proof for $\g\neq 0$ can be found in
\cite[Theorem I-3.1.4]{aghh}. From Weyl's theorem (see \cite[Theorem
IV-5.35]{k}), the essential spectrum of both operators $\Lug$ and
$\Ldg$ is $[\w,+\infty)$. Since both operators are bounded from
below, there can be only finitely many isolated eigenvalues (of
finite multiplicity) in $(-\infty,\w')$ for any $\w'<\w$. Then (i)
and (iii) follow easily.

Next, we consider (ii).
Since $\fwg$ satisfies $\Ldg\fwg=0$ and $\fwg>0$, the first eigenvalue of $\Ldg$ is $0$ and the rest of the spectrum is positive . This is classical for $\g=0$ and can be easily proved for $\g\neq 0$, see \cite[Chapter 2, Section 2.3, Paragraph 3]{bs}. Thus to ensure that the kernel of $\Hwg$ is reduced to $\mbox{\rm span}\{i\fwg\}$ it is enough to prove that the kernel of $\Lug$ is $\{0\}$. It is equivalent to prove that $0$ is the unique solution of
\begin{equation}
\Lug u=0,\;u\in D(\Lug).\label{eq1ev}
\end{equation}
To be more precise, the solutions of (\ref{eq1ev}) satisfy
\begin{eqnarray}
&&u\in H^2(\R\setminus\{0\})\cap\hu,\label{eq3ev}\\
&&-\dxx u+\w u-p\fwg^{p-1} u=0,\label{eq4ev}\\
&&\dx u(0+)-\dx u(0-)=-\g u(0).\label{eq5ev}
\end{eqnarray}
Consider first (\ref{eq4ev}) on $(0,+\infty)$. If we look at (\ref{snls}) only on $(0,+\infty)$, we see that $\fwg$ satisfies
\begin{equation}
-\dxx \fwg+\w \fwg-\fwg^{p}=0\mbox{ on }(0,+\infty).\label{eq2ev}
\end{equation}
If we differentiate (\ref{eq2ev}) with respect to $x$
(which is possible because $\fwg$ is smooth on $(0,+\infty)$),
we see that $\dx \fwg$ satisfies (\ref{eq4ev}) on $(0,+\infty)$.
Since we look for solutions in $\ld$ (in fact solutions going to $0$ at infinity),
it is standard that every solution of (\ref{eq4ev}) in $(0,+\infty)$
is of the form $\m \dx \fwg$, $\m\in\R$
(see, for example, \cite[Chapter 2, Theorem 3.3]{bs}).
A similar argument can be applied to (\ref{eq4ev}) on $(-\infty,0)$,
thus every solution of (\ref{eq4ev}) in $(-\infty,0)$ is
of the form $\n \dx \fwg$, $\n\in\R$.

Now, let $u$ be a solution of (\ref{eq3ev})-(\ref{eq5ev}).
Then there exists $\m\in\R$ and $\n\in\R$ such that
\begin{eqnarray*}
u=\n \dx \fwg&\mbox{ on }&(-\infty,0),\\
u=\m \dx \fwg&\mbox{ on }&(0,+\infty).
\end{eqnarray*}
Since $u\in\hu$, $u$ is continuous at $0$,
thus we must have $\m=-\n$, that is $u$ is of the form
\begin{eqnarray*}
&&u=-\m \dx \fwg\mbox{ on }(-\infty,0),\\
&&u=\phantom{-}\m \dx \fwg\mbox{ on }(0,+\infty),\\
&&u(0)=-\m \dx \fwg(0-)=\m \dx \fwg(0+)=\frac{-\m}{2}\g\fwg(0).
\end{eqnarray*}

Furthermore, $u$ should satisfies the jump condition (\ref{eq5ev}).
Since $\fwg$ satisfies
$$
\dxx\fwg(0-)=\dxx\fwg(0+)=\w\fwg(0)-\fwg^{p}(0),
$$
if we suppose $\m\ne 0$ then (\ref{eq5ev}) reduces to
$$
\fwg^{p-1}(0)=\frac{4\w-\g^2}{4}.
$$
But from (\ref{eqexplicitfwg}) we know that
$$
\fwg^{p-1}(0)=\frac{p+1}{8}(4\w-\g^2).
$$
It is a contradiction, therefore $\m=0$. In conclusion, $u\equiv
0$ on $\R$, and the lemma is proved.
\vspace{3mm}

{\it Proof of Lemma \ref{lemanalyticityLug}.}
We recall that $\Lg$ is defined with the help of a bilinear
form $\Bugw$ (see (\ref{eqBugBdg})).
To prove the holomorphicity of $(\Lg)$
it is enough to prove that $(\Bugw)$ is bounded from below and closed,
and that for any $v\in\hu$
the function $\Bugw(v):\g\mapsto \Bugw(v,v)$ is holomorphic
(see \cite[Theorem VII-4.2]{k}).
It is clear that $\Bugw$ is bounded from below and closed
on the same domain $\hu$ for all $\g$,
thus we just have to check the holomorphicity
of $\Bugw(v):\g\mapsto \Bugw(v,v)$ for any $v\in\hu$.
We recall the decomposition of $\Bugw$
into $\Buug$ and $\Budgw$ (see (\ref{eqBuugBudg})).
We see that $\Buug(v)$ is clearly holomorphic in $\g$.
            From the explicit form of $\fg$ (see (\ref{eqexplicitfwg}))
it is clear that $\g\mapsto\fg^{p-1}(x)$ is holomorphic in $\g$
for any $x\in\R$.
It then also follows that $\g\mapsto\Budgw(v)$ is holomorphic.
\hfill\qed

\begin{rmq}
There exists another way to show that $(\Lg)$ is a real-holomorphic family with respect to $\g\in \R$.
We can use the explicit resolvent formula in \cite{aghh},
$$(T_1-k^2)^{-1}=(-\dxx-k^2)^{-1}+2\g k(-i\g+2k)^{-1}
(\overline{G_k(\cdot)},\cdot)G_k(\cdot),$$
where $k^2 \in \rho(T_1)$, $\Im k>0$, $G_k(x)=(i/2k)e^{ik|x|}$,
to verify the holomorphicity.
\end{rmq}

\section{Proof of Lemma \ref{lem:impaire}}\label{ap:impaire}

First, we indicate how the extension of $f$ to $(-\infty,+\infty)$ can be done.
We see by the proof in \cite[Theorem XII.8]{rs} that
the functions $f(\g)$ and $\l(\g)$ defined
in Lemma \ref{lemanalyticity} exist,
are holomorphic and represent an eigenvector
and an eigenvalue for all $\g \in \R$,
since $(\Lg)$ is a real-holomorphic family in $\g \in \R$.
Namely we can repeat the argument of Lemma \ref{lemanalyticity}
at each point $\g $ and on each neighborhood of $\g$.
This is possible because the set $\{(\g, \l);~\g \in \R,~ \l \in \rho(\Lg)\}$
is open and the function $(\Lg-\l)^{-1}$ defined
on this set is a holomorphic function
of two variables (\cite[Theorem XII.7]{rs}).

Secondly, as it was observed in \cite{fj,fsw},
the eigenvectors of $\Lg$ are even or odd.
Indeed, let $\xi$ be an eigenvalue of $\Lg$ with eigenvector $v\in D(\Lg)$.
Then clearly $\tilde{v}$ with $\tilde{v}(x)=v(-x)$
is also an eigenvector associated to $\xi$. In particular, $v$ and $\tilde{v}$ satisfy both
$$
-\dxx v+(\w-\xi) v-p\fg^{p-1}v=0\mbox{ on }[0,+\infty),
$$
thus there exists $\eta\in\R$ such that $v=\eta\tilde{v}$ on $[0,+\infty)$ (this is standard, see, for example, \cite[Chapter 2, Theorem 3.3]{bs}). If $v(0)\neq 0$, it is immediate that $\eta=1$.
If $v(0)=0$, then $\dx v(0+)\neq 0$ (otherwise the Cauchy-Lipschitz Theorem leads to $v\equiv 0$),
and it is also immediate that $\eta=-1$.
Arguing in a same way on $(-\infty,0]$,  we conclude that $v$ is even or odd,
and in particular $v$ is even if and only if $v(0)\neq 0$.

Finally, we prove the last statement only for the case $\g<0$
since the case $\g>0$ is similar. We remark that $\dx\fo$ is odd.
Since $\lim_{\g \to 0}\psld{f(\g)}{\dx\fo}=\nldd{\dx\fo}\neq 0$,
we have $\psld{f(\g)}{\dx\fo}\neq 0$ for $\g$ close to $0$,
thus $f(\g)$ cannot be even, and therefore $f(\g)$ is odd.
Let $\tilde{\g}_{\infty}$ be
$$
\tilde{\g}_{\infty}
=\inf\{\tilde{\g}<0;~f(\g)\mbox{ is odd for any }\g\in(\tilde{\g},0] \}.
$$
We suppose that $\tilde{\g}_{\infty} >-\infty$. If
$f(\tilde{\g}_{\infty})$ is odd, by continuity in $\g$ of $f(\g)$
with $L^2$ value, there exists $\varepsilon>0$ such that
$f(\tilde{\g}_{\infty}-\varepsilon)$ is odd which is a
contradiction with the definition of $\tilde{\g}_{\infty}$, thus
$f(\tilde{\g}_{\infty})$ is even. Now, $f(\tilde{\g}_{\infty})$ is
the limit of odd functions, thus is odd. The only possibility to
have $f(\tilde{\g}_{\infty})$ both even and odd is
$f(\tilde{\g}_{\infty})\equiv 0$, which is impossible because
$f(\tilde{\g}_{\infty})$ is an eigenvector.
\hfill\qed

\section{Proof of Proposition \ref{thmvirial}} \label{secvirial}
\renewcommand{\a}{a}

For $\a\in\N\setminus\{ 0 \}$, we define $\Va(x)=\g\a e^{-\pi\a^2 x^2}$. It is clear that $\intr\Va(x)=\g$ and $\Va\goestoweak\g\d$ weak-$\star$ in $\hmu$ when $\a\goesto +\infty$.

We begin by the construction of approximate solutions : for
\begin{equation}\label{nlsa}
\left\{
\begin{array}{rcl}
i\dt u & = & -\dxx u -\Va u -|u|^{p-1}u,\\
u(0) & = & \uo,
\end{array}
\right.
\end{equation}
and thanks to \cite[Proposition 6.4.1]{c}, for every $\a\in\N\setminus\{ 0 \}$ there exists $\Ta>0$ and a unique maximal solution $\ua\in\mathcal{C}([0,\Ta),\hu)\cap\mathcal{C}^1([0,\Ta),\hmu)$ of (\ref{nlsa}) which satisfies for all $t\in[0,\Ta)$
\begin{eqnarray}
\Ea(\ua(t))&=&\Ea(\uo),\label{conservationenergya}\\
\nld{\ua(t)}&=&\nld{\uo},\label{conservationchargea}
\end{eqnarray}
where
$\displaystyle
\Ea(v)  =  \frac{1}{2}\nldd{\dx v}-\frac{1}{2}\intr\Va|v|^2dx-\frac{1}{p+1}\nlpsps{v}.
$
Moreover, the function
$
\fa:t\mapsto\intr x^2|\ua(t,x)|^2 dx
$
is $\mathcal{C}^2$ by \cite[Proposition 6.4.2]{c}, and
\begin{eqnarray}
\dt \fa  & = & 4\Im\intr \overline{\ua}x\dx\ua dx,\label{eq1}\\
\dtt \fa & = & 8\Ka(\ua)\label{eq2}
\end{eqnarray}
where $\Ka$ is defined for $v\in\hu$ by
$$
\Ka(v)= \nldd{\dx v}+\frac{1}{2}\intr x(\dx\Va)|v|^2 dx
-\frac{p-1}{2(p+1)}\nlpsps{v}.
$$

Then, we find estimates on $(\ua)$. Let $M\geq\nhu{\uo}$ (an exact
value of $M$ will be precise later). We define
\begin{equation}\label{ta}
\ta=\sup\{t>0; \nhu{\ua(s)}\leq 2M\mbox{ for all }s\in[0,t)\}.
\end{equation}
Since $\ua$ satisfies (\ref{nlsa}), we have
$$
\displaystyle\sup_{\a\in\N\setminus\{ 0 \}}\| \dt\ua \|_{L^\infty([0,\ta),\hmu)}\leq C,
$$
and thus for all $t\in[0,\ta)$ and for all $\a\in\N\setminus\{ 0 \}$ we get
\begin{equation}\label{eq5}
\nldd{\ua(t)-\uo}=2\int_0^t \psld{\ua(s)-\uo}{\dt\ua(s)}ds\leq C t
\end{equation}
where $C$ depends only on $M$. Now we have
$$
\frac{1}{p+1}(\nlpsps{\ua}-\nlpsps{\uo})=\int_0^1 \! \intr  (\ua-\uo)|s\ua+(1-s)\uo|^p dx\,    ds
$$
which combined with H\"older inequality, Sobolev embeddings, (\ref{ta}) and (\ref{eq5}) gives
\begin{equation}\label{eq5b}
\frac{1}{p+1}(\nlpsps{\ua}-\nlpsps{\uo})\leq Ct^{1/2}.
\end{equation}
Moreover, using (\ref{ta}), Sobolev embeddings, Gagliardo-Nirenberg inequality and (\ref{eq5}) we obtain
\begin{equation}\label{eq5c}
\left| \intr\Va(|\ua|^2-|\uo|^2) \right|\leq Ct^{1/4}.
\end{equation}
Combining (\ref{conservationenergya}), (\ref{conservationchargea}), (\ref{eq5b}) and (\ref{eq5c}) leads to
$$
\nhud{\ua(t)}\leq M^2+C(t^{1/4}+t^{1/2})\mbox{ for all $t\in[0,\ta)$ and for all $\a\in\N\setminus\{ 0 \}$},
$$
and choosing $T_M$ (depending only on $M$) such that $C(T_M^{1/4}+T_M^{1/2})=3M^2$ we obtain for all $\a\in\N\setminus\{ 0 \}$ the estimates
\begin{equation}\label{eq6}
\begin{array}{l}
\| \ua \|_{L^{\infty}([0,T_M),\hu)}\leq 2M,\\
\| \dt\ua \|_{L^{\infty}([0,T_M),\hmu)}\leq C.\\
\end{array}
\end{equation}
In particular, it follows from (\ref{eq6}) that $T_M\leq\ta$ for all $\a\in\N\setminus\{ 0 \}.$

Now we can pass to the limit : thanks to (\ref{eq6}) there exists $u\in L^{\infty}([0,T_M),\hu)$ such that for all $t\in[0,T_M)$ we have
\begin{equation}\label{eq7}
\ua(t)\goestoweak u(t) \mbox{ weakly in }\hu\mbox{ when }\a\goesto +\infty,
\end{equation}
which immediately induces that when $\a\goesto +\infty$,
\begin{equation}\label{eq8}
|\ua(t)|^{p-1}\ua(t)\goestoweak |u(t)|^{p-1}u(t) \mbox{ weakly in }\hmu.
\end{equation}
In particular, thanks to Sobolev embeddings, we have
$$
\ua(t,x)\goesto u(t,x)\mbox{ a.e. and uniformly on the compact sets of }\R,
$$
and it is not hard to see that it permit to show
\begin{equation}\label{eq9}
\Va\ua\goestoweak u\g\d \mbox{ weak-$\star$ in }\hmu.
\end{equation}
Since $\ua$ satisfies (\ref{nlsa}), it follows from (\ref{eq7}), (\ref{eq8}) and (\ref{eq9}) that $u$ satisfies (\ref{nls}). Finally, by (\ref{conservationcharge}) and (\ref{conservationchargea}), we have
$$
\ua\goesto u\mbox{ in }\mathcal{C}([0,T_M),\ld),
$$
thus, from Gagliardo-Nirenberg inequality and (\ref{eq6}),  we have
$$
\ua\goesto u\mbox{ in }\mathcal{C}([0,T_M),\lps),
$$
and by (\ref{conservationenergy}) and (\ref{conservationenergya}) it follows that
\begin{equation}\label{eq11}
\ua\goesto u\mbox{ in }\mathcal{C}([0,T_M),\hu).
\end{equation}

We have to prove that the time interval $[0,T_M)$ can be extended as large as we need. Let $0<T<\Tuo$ and
$$
M=\sup\{ \nhu{u(t)},t\in[0,T] \}.
$$
If $T_M\geq T$, there is nothing left to do, thus we suppose
$T_M<T$. From (\ref{eq11}) we have $\nhu{\ua(T_M)}\leq M$ for $\a$
large enough. By performing a shift of time of length $T_M$ in
(\ref{nls}) and (\ref{nlsa}) and repeating the first steps of the
proof we obtain
$$
\ua\goesto u\mbox{ in }\mathcal{C}([T_M,2T_M),\hu).
$$
Now we reiterate this procedure a finite number of times until we covered the interval $[0,T]$ to obtain
\begin{equation}\label{eq12}
\ua\goesto u\mbox{ in }\mathcal{C}([0,T],\hu).
\end{equation}

To conclude, we remark that (\ref{eqvirial1}) follows from the
same proof than \cite[Lemma 6.4.3]{c} (computing with
$\nldd{e^{\varepsilon|x|^2}xu(t)}$ and passing to the limit
$\varepsilon\goesto 0$), thus we have
\begin{equation}\label{eq10}
\nldd{xu(t)}=\nldd{x\uo}+4\int_0^t\Im\intr\overline{u(s)}x\dx u(s)dxds.
\end{equation}
From (\ref{eq1}), Cauchy-Schwartz inequality and (\ref{eq6}) we have
$$
\dt \left(\nldd{x\ua(t)}\right)\leq C\nld{x\ua(t)},
$$
which implies that
$$
\nld{x\ua(t)}\leq\nld{x\uo}+Ct.
$$
Since in addition we have
$$
x\ua(t,x)\goesto xu(t,x)\mbox{ a.e.},
$$
we infer that
$$
x\ua(t,x)\goestoweak xu(t,x)\mbox{ weakly in }\ld.
$$
Recalling that
$$
\dx\ua\goesto\dx u\mbox{ strongly in }\ld
$$
we can pass to the limit in (\ref{eq10}) to have
$$
\nld{x\ua(t)}\goesto\nld{xu(t)}.
$$
On the other side, since we have (\ref{eq2}) and (\ref{eq12}), we get (\ref{eqvirial2}).

\begin{rmq}
Our method of approximation is inspired of the one developed in \cite{cw} by
Cazenave and Weissler to prove the local well-posedness of the Cauchy problem
for nonlinear Schr\"{o}dinger equations. Actually, slight modifications in our
proof of Proposition~\ref{thmvirial} would permit to give an alternative proof of
Proposition \ref{wpd}.
\end{rmq}

\end{document}